\documentclass[journal]{IEEEtran}

\IEEEoverridecommandlockouts
\usepackage{amsthm}
\usepackage{amsfonts}
\usepackage{amssymb}
\usepackage{mathrsfs}
\usepackage{mathtools}
\usepackage{adjustbox}
\usepackage[english]{babel}
\usepackage{float}
\usepackage{graphicx}
\usepackage{epstopdf}
\usepackage{threeparttable}
\usepackage{amsmath}
\usepackage{color}
\usepackage{multirow}
\usepackage{algpseudocode}
\usepackage{algorithm}
\usepackage{algorithmicx}
\usepackage{balance}
\let\labelindent\relax
\usepackage[inline]{enumitem}
\usepackage[caption=false,font=footnotesize]{subfig}

\usepackage{cite}

\usepackage{hyperref}
\hypersetup{colorlinks,allcolors=black}
\usepackage{graphicx}
\usepackage{textcomp}
\usepackage{xcolor}
\def\BibTeX{{\rm B\kern-.05em{\sc i\kern-.025em b}\kern-.08em
    T\kern-.1667em\lower.7ex\hbox{E}\kern-.125emX}}
\DeclareMathAlphabet\mathbfcal{OMS}{cmsy}{b}{n}
\newtheorem{theo}{Theorem}

\newtheorem{propo}{Proposition}

\theoremstyle{definition}

\DeclareMathOperator*{\argmax}{arg\,max}
\title{CoFAR Clutter Estimation using Covariance-Free \\Bayesian Learning}

\author{Kunwar~Pritiraj~Rajput,~\IEEEmembership{Member,~IEEE,}
        Bhavani~Shankar~M.~R.,~\IEEEmembership{Senior Member,~IEEE,}
        Kumar~Vijay~Mishra,~\IEEEmembership{Senior Member,~IEEE,} Muralidhar~Rangaswamy,~\IEEEmembership{Fellow,~IEEE}, and 
        Bjorn~Ottersten,~\IEEEmembership{Fellow,~IEEE}    
\thanks{K. P. R., M. R. B. S., and B. O. are with the interdisciplinary center of security reliability and trust (SnT), University of Luxembourg, 1855, Luxembourg (e-mail: \{kunwar.rajput, bhavani.shankar, bjorn.ottersten\}@uni.lu). K. V. M. is with the United States DEVCOM Army Research Laboratory, Adelphi, Maryland, USA (e-mail: kvm@ieee.org). M. R. is with the United States Air Force Research Laboratory, Wright-Patterson Air Force Base, Ohio, USA  (email:muralidhar.rangaswamy@us.af.mil). The conference precursor to this work \cite{10149624} was published in the 2023 IEEE Radar Conference (RadarConf).}}
 
\begin{document}

\maketitle

\begin{abstract}
A cognitive fully adaptive radar (CoFAR) adapts its behavior on its own within a short period of time in response to changes in the target environment. For the CoFAR to function properly, it is critical to understand its operating environment through estimation of the clutter channel impulse response (CCIR). In general, CCIR is sparse but prior works either ignore it or estimate the CCIR by imposing sparsity as an explicit constraint in their optimization problem. In this paper, contrary to these studies, we develop covariance-free Bayesian learning (CoFBL) techniques for estimating sparse CCIR in a CoFAR system. In particular, we consider a multiple measurement vector scenario and estimate a simultaneously sparse (row sparse) CCIR matrix. Our CoFBL framework reduces the complexity of conventional sparse Bayesian learning through the use of the diagonal element estimation rule and conjugate gradient descent algorithm. We show that the framework is applicable to various forms of CCIR sparsity models: group, joint, and joint-cum-group. We evaluate our method through numerical experiments on a data set generated using RFView®, a high-fidelity modeling and simulation tool. We derive Bayesian Cram\'{e}r-Rao bounds for the various considered scenarios to benchmark the performance of our algorithms. Our results demonstrate that the proposed CoFBL-based approaches perform better than the existing popular approaches such as multiple focal underdetermined system solver and simultaneous orthogonal matching pursuit.
\end{abstract}


\section{Introduction}
A cognitive fully adaptive radar (CoFAR) accumulates the knowledge of the target environment and subsequently, optimizes its transmit and receive functions adaptively for enhanced performance \cite{Haykin2006, Bergin2005, Guerci2010, Guerci2014}. The purpose of adaptive processing may include desired results such as target identification \cite{6875604} and tracking \cite{6875605}. Conventional radars use a feed-forward mechanism, which transfers raw detected data to a higher-level processor for more sophisticated functions. The FAR system, on the other hand, uses a feedback mechanism and optimization criteria to enhance sensing in the following scans. In the case of a multi-function radar that performs both detection and imaging, the optimization may have additional goals such as tracking, imaging, and classification \cite{8798987}. Hardware prototypes that address certain FAR goals have also been exhibited. For instance, \cite{7944305} designed a FAR system for target tracking in a non-linear range measurement model. In this paper, we focus on clutter channel impulse response (CCIR) estimation in the context of CoFAR.

To effectively analyze the target environment, it is important to estimate CCIR \cite{1456694,1054205}, which is made up of undesirable echoes including ground reflections, atmospheric effects, and electromagnetic interference \cite{wetzel2008sea,bringi2024advances2,billingsley2002low,ward2006sea,kulemin2003millimeter,musha1994models}. These factors hinder radar performance by reducing detection and estimation capabilities \cite{4200703}. Kay in \cite{4200703} proposed an optimal signal design for a Gaussian single target in the presence of clutter which is also modeled as Gaussian random process (GRP). In space time adaptive processing (STAP) the traditional clutter model establishes a non-linear relationship between the signal-dependent covariance and the transmit radar waveform, making it difficult to accurately estimate the signal-dependent clutter covariance matrix. However, the multi-input multi-output (MIMO) clutter model proposed in \cite{Guerci2016} employs a ``stochastic transfer function" to overcome this issue. This transfer function is derived from a physical scattering model and may be viewed as a stochastic Green's function \cite{greenberg1971applications}. Although the complex clutter returns are influenced by the transmit signal, the stochastic transfer function is signal-independent. This approach allows for joint optimization of the transmit waveform and receiver filters because the received signal is the convolution of the transmit waveform with the target and CCIRs. 

For the waveform design problem,
\cite{Guerci2016} presents an analytical solution that requires estimation from representative training data in practice. However, the accuracy of the channel transfer function estimates heavily influences the performance of this method. To address this issue, a low-complexity frequency-domain technique was proposed in \cite{Gogineni2019} for estimating the channel matrix. This estimate is then incorporated into an optimization problem thereby allowing for the simultaneous design of the radar transmit-receive functions to achieve the CoFAR goal. Although this technique is easily implemented in real-time hardware, accurately estimating the CIR at low signal-to-noise ratio (SNR) levels requires probing signals with larger pulse-widths. This calls for a significant allocation of CoFAR resources thereby rendering this approach undesirable.

{CoFAR performance is critically dependent on the knowledge of the CCIR \cite{ Kang2020,kang2021,9764307,gogineni2022high}. Devoting CoFAR resources in order to gain that knowledge takes away from the radar dwell time for the primary surveillance function. Therefore, one has to exercise caution in terms of how often and how to probe the environment. Thus, exploiting structure inherent to the CCIR such as sparsity or constancy along certain band like constrained CCIR \cite{kang2021,9764307,gogineni2022high} helps in gaining that knowledge with less CoFAR resources. This affords a significant performance improvement by cutting down on the radar dwell time.} To efficiently use CoFAR resources, it would be beneficial to take advantage of inherent structural details or relationships between nearby channel transfer functions in the estimation problem. To this end, \cite{Kang2020,kang2021} exploit the fact that CIRs of neighboring pulses are often strongly correlated. This allows for the use of a constrained maximum likelihood (ML) estimate employing a cosine similarity constraint. However, the resulting optimization problems are semidefinite program (SDP) and second order cone program (SOCP) which entails high computational complexity. In \cite{9764307}, a constrained CIR estimation (CCIRE) was introduced which utilized both a cosine similarity constraint and sparsity in CIR. However, implementing this optimization method  would also results in a computationally expensive SDP. Taking this paradigm forward, \cite{gogineni2022high} developed a framework that exploits the cosine similarity and conducts experiments on the RFView®-generated real data sets. 

An alternative approach to tackling the estimation problem involves leveraging a sparse signal recovery technique, such as Basis Pursuit \cite{chen2001atomic} or the FOCal Underdetermined System Solver (FOCUSS) \cite{gorodnitsky1997sparse}, to estimate the sparse CCIR. However, these methods come with fixed priors and are susceptible to structural and convergence errors. Moreover, they necessitate the tuning parameter $\lambda$ to be determined through another method or set based on prior knowledge. Any disparity between the assumed prior and the data may lead to a substantial loss in performance. Structural errors escalate when the global minimum of the cost function is not the sparsest solution, and convergence errors arise when the method converges to a sub-optimal local minimum. In contrast, the sparse Bayesian learning (SBL) framework \cite{1315936} has garnered considerable research attention as a tool for estimating the unknown sparse quantity while offering superior performance guarantees. Its application for simultaneously estimating sparse parameters was further extended in \cite{4244754}. SBL has found application in various fields, including biomedical signal processing \cite{Saha2015SparseBL} and array processing \cite{hu2016source,8529199}.

The parameterized prior in SBL does not require to tweak unknown regularization parameter. Moreover, it uses a tractable expectation maximization (EM) framework, which yields maximally sparse solution \cite{1315936}. In the absence of noise, the global minimum occurs at the sparsest solution and there are no structural errors. For noisy signals, SBL performs reasonably well since the SBL cost function also has fewer local minima than the FOCUSS objective function, making it less susceptible to convergence errors \cite{1315936}. 

{\begin{table*}[t]
\centering
 \caption{Canvas of the contributions of the current work vis-a-vis the state-of-the-art.}
\label{Literature}
\begin{adjustbox}{max width=2\columnwidth}
\begin{threeparttable}
\begin{tabular}{||l|c|c|c|c|c|c|c|c|c|c|c|c|c|c|c|c|c|c|c|c|c|}
\hline
q.v. & Radar  & Antenna&Measurements& Estimation & Clutter& Sparsity &  Priors & Simulation \\
 & type & geometry& & method & model & & & tool\\
\hline
\hline
[1] & CoFAR &SIMO & SMV  & Bayesian & Green's function & Standard & Not required & RFView\\
\hline
[3]  & CoFAR & MIMO  & SMV & Least Square & Delay-Doppler & No & N/A & Simulated \\
\hline
[5]  & CoFAR &  MIMO  & SMV  & STAP & Covariance based& No& N/A& Simulated\\
\hline
 [17] & Conventional & SISO & SMV  & Bayesian & GRP & No & N/A & Simulated\\
\hline
[20]  & CoFAR & MIMO & SMV   & Frequency domain & Green's function & No & N/A & RFView\\
\hline
[21]  & CoFAR & MIMO & SMV   & SDP & Green's function & No & N/A & RFView\\
\hline
[22]  & CoFAR & MIMO & SMV   & SOCP & Green's function & No & N/A & RFView\\
\hline
[23]  & CoFAR & MIMO &  SMV & SDP & Green's function & Standard & Yes & RFView\\
\hline
 \textbf{This paper}  & CoFAR & MIMO  & MMV & Bayesian& Green's function & Standard, joint, group, & Not required & RFView\\
 & & & &  &  &  joint-cum-group & & \\
 \hline
\end{tabular}\label{Table_LR}
 \begin{tablenotes}
            \footnotesize
            \item[1] N/A: Not applicable.
        \end{tablenotes}
    \end{threeparttable}
\end{adjustbox}
\end{table*}} 

The EM algorithm poses increased computational complexity due to the necessity of inverting a posterior covariance matrix. To mitigate this complexity, alternative algorithms based on Approximate Message Passing (AMP) \cite{7457279} and Gaussian Belief Propagation (GBP) \cite{5382564} can be employed. These approaches aim to reduce computational demands by conducting $T_{amp}$ and $T_{gbp}$ inner steps within each E-Step, estimating means and variances of the sparse parameters without resorting to matrix inversion. However, it's worth noting that AMP and GBP may encounter convergence issues, particularly in cases involving the estimation of variances \cite{6872556}. An instance is when the dictionary fails to satisfy zero-mean and sub-Gaussian (i.e., distribution with a strong tail decay) criteria \cite{8074806}.

A different approach involves incorporating variational inference, aiming to approximate the actual posterior using a more manageable surrogate \cite{bishop2000variational,5946762,5946760}. This technique facilitates inverse-free Sparse Bayesian Learning (SBL) inference \cite{8907445,7894261}. However, variational inference methods optimize a lower bound on the likelihood function rather than the true objective, potentially leading to convergence on a sub-optimal solution for the sparse quantity in question. In the context of this study, we also tackle the challenge of reducing the computational complexity of the Expectation-Maximization (EM) algorithm by eliminating the need for inverting the posterior covariance matrix in each EM iteration.

\subsection{Our Contributions} 
Preliminary results of our work appeared in our conference publication \cite{10149624}, where we proposed a sparse CCIR vector estimation technique using conventional SBL \cite{1315936} for a single-input multiple-output (SIMO) CoFAR system. This required only a single measurement vector model. 
In this paper, our main contributions are:
\begin{itemize}
\item {\textbf{MIMO CoFAR:} We consider a multiple measurement vector (MMV) scenario and derive the SBL-based framework for the CCIR matrix estimation for a MIMO CoFAR. The proposed algorithm exploits the simultaneous or row sparsity in the range domain using the MMV framework. Prior works  \cite{10149624,Bergin2005,Guerci2014,Gogineni2019,Kang2020,kang2021,9764307} relied on single measurement vector (SMV) models.}
\item \textbf{Reduced computations in EM algorithm:} We develop the covariance-free Bayesian learning (CoFBL) framework \cite{9807393} for the MMV model to substantially reduce the computational requirement of the E-step which requires the inversion of the covariance matrix. This requires computation of the diagonal elements of the posterior covariance matrix along with the posterior mean. Conventionally, this entails the inversion of the posterior covariance matrix but we avoid it using the diagonal estimation rule \cite{bekas2007estimator} together with conjugate gradient descent (CGD) algorithm \cite{hestenes1952methods} to solve linear system of equations. 
\item {\textbf{Joint, group, and joint-cum-group sparse models:} We leverage upon the various sparse models of the underlying channel and develop covariance-free group (CoFGBL), joint (CoFJBL) and joint-cum-group Bayesian learning (CoFJGBL) techniques. This requires the computation of fewer hyperparameters and still yields a better estimation performance.}
\item \textbf{Comparisons with realistic datasets:} Previous related works \cite{4200703,1593336,825822} rely on simplified or highly theoretical models for datasets. We employ the high-fidelity modeling and simulation software RFView® to realistically evaluate our methods against the state-of-the-art.
\item \textbf{Bayesian error bounds:} To benchmark the results, we derive Bayesian Cram\'{e}r-Rao bounds (BCRBs) for various algorithms proposed in this work. We build upon the deterministic BCRB and specialize it to particular cases of interest for radar channel estimation. These scenarios rely on the identification of structures not exploited hitherto. 

\end{itemize}  
{Table \ref{Table_LR} summarizes our contributions with respect to state-of-the-art.} 

\subsection{Organization and Notation}
The rest of the paper is structured as follows. In the next section, we introduce the MMV system model for MIMO CoFAR. Section III describes the BL-based CCIR estimation algorithm using the conventional sparse BL (SBL) algorithm. In Section IV, we develop the low-complexity CoFBL framework for CCIR estimation. Then, we expand this in Section V to the proposed CoFGBL, CoFJBL, CoFJGBL methods. In Section VI, we derive the BCRBs for the different scenarios considered in this paper. We validate our models and methods through numerical experiments in Section VII and conclude in Section VIII followed by appendices.

Lowercase and uppercase bold letters represent vectors and matrices, respectively. The superscripts $T$ and $H$ denote transpose and Hermitian (conjugate transpose) operations, respectively. The statistical expectation and trace operator are defined as $\mathbb{E}[.]$ and $\mathrm{Tr}[.]$, respectively. The expression $\mathbf{x} \sim \mathcal{CN}\left(\boldsymbol{\mu},\mathbf{\Sigma}\right)$ represents a complex vector $\mathbf{x}$ with mean $\boldsymbol{\mu}$ and covariance matrix $\mathbf{\Sigma}$. The $n$-th element of a vector $\mathbf{x}$ is denoted by $\mathbf{x}(n)$, and the $(n,n)$-th element of a matrix $\mathbf{X}$ is denoted as $\mathbf{X}(n,n)$. The $i$-th row and $j$-th column of any matrix $\mathbf{X}$ are denoted as $\mathbf{X}_{i.}$ and $\mathbf{X}_{.j}$, respectively. The operation $\mathbf{vec}(.)$ converts an $n \times n$ matrix $\mathbf{X}$ into an $n^2 \times 1$ vector by performing column-wise stacking. The expression $\mathbf{X}=\textrm{diag}[x_1,x_2,\hdots,x_N] \in \mathbb{C}^{N \times N}$ represents a diagonal matrix with elements $x_i \in \mathbb{C}$, $i=1,2,\hdots,N$ on its main diagonal. Finally, $\ast$ and $\odot$ ($\oslash$) denote the convolution, element-wise multiplication (division), respectively.

\section{System Model}
Consider a MIMO radar with $N$ transmit and $M$ receive antenna elements arranged in an uniform linear array (ULA) configuration with inter-element spacing $d_T$ and $d_R$, respectively. The inter-element spacing of the
transmit array admits values higher than half-wavelength
without suffering from ambiguity at the receiver \cite{chen2008mimo}. Therefore, we consider $d_R=\lambda/2$ so that $d_T=M\lambda/2$. The total aperture of the transmit and receive arrays is $Z_t$ and $Z_r$, respectively. Denote $Z=Z_t+Z_r$. Then, the locations of $n$-th transmit and $m$-th receive antennas on the x-axis are $\frac{Znd_T}{2}$ and $\frac{Zmd_R}{2}$, respectively. At each transmitter $n$, the transmit radar signal $x_n(t)$ is 
\begin{equation}
\label{eq:LFM}
    x_n(t) = s_n(t)e^{j2 \pi f_c t},
\end{equation}
where $f_c$ is carrier frequency and $s_n(t)$ is a linear frequency modulated (LFM) baseband waveform
	\begin{align}
	s_n(t)&=A_n e^{\mathrm{j}2\pi  \beta_n t^2}\textrm{rect}\left(\frac{t}{T}\right),\;0<t<T,
	\end{align} 
where $A_n$ is the signal amplitude, $\beta_n = \frac{B_n}{2T}$ is the ramp rate, $B_n$ is the non-overlapping signal bandwidth of $n$-th transmit signal, and
	\begin{equation}
	\textrm{rect}(t)= 
\begin{cases} 
    1, \phantom{1}\phantom{1} 0\leq t < 1\\
    0, \phantom{1}\phantom{1}\text{otherwise},
\end{cases}
	\end{equation} 
is the rectangular pulse. The pulse repetition interval (PRI) is denoted by  $T$. The LFM for the transmit waveform has a wide sweep bandwidth and is appropriate for high-resolution sensing of the channel.

Assuming far-field point targets, the $N\times 1$ transmit steering vector is
\begin{align}
\label{eq:pa_txvec}
\mathbf{c}(\theta) &= [c_1(\theta), \cdots, c_N(\theta)]^T \nonumber\\
&= [1, e^{-j\frac{2\pi d_T}{\lambda}\theta}, \cdots, e^{-j\frac{2\pi (N-1)d_T}{\lambda}\theta}]^T,
\end{align}
Then, the transmit signal for $k$-th pulse at the angular location $\theta$ is
\begin{align}
\sum_{n=1}^{N-1} e^{-j\frac{2\pi nd_T}{\lambda}\theta} x_n(t-kT) \triangleq \mathbf{c}^T(\theta) \mathbf{x}(t-kT),
\end{align}
where $\mathbf{x}(t) = \left[x_1(t), \ldots, x_N(t)\right]^T \in \mathbb{C}^{N \times 1}$. 

 Assume $R$ scatterers on the scene, each of which is characterized by the complex reflectivity $\alpha_{r} \in \mathbb{C}$ that is assumed constant over the array for far-field targets; time-delay $\tau_{m,n,r} \in \mathbb{R}$ corresponding to the $m$-th receive and $n$-th transmit antennas, direction-of-departure (DoD) $\theta_{n,r} = \sin\zeta_{n,r}\in \mathbb{R}$, and direction-of-arrival (DoA) $\phi_{m,r} = \sin\zeta_{m,r}\in \mathbb{R}$, where $\zeta_{n,r}$ and $\zeta_{m,r}$ are the angular locations corresponding to $n$-th transmit and $m$-th receive antenna for $r$-th clutter scatterer. The corresponding CCIR is
\begin{align}
    h_{m,n}&(\tau, \theta,\phi) \nonumber\\ &= \sum_{r=1}^{R} \alpha_{r} \delta(\tau-\tau_{m,n,r})\delta(\theta-\theta_{n,r})\delta(\phi-\phi_{m,r}).
\end{align}
For collocated arrays, the complex reflectivity is also constant over both transmit and receive arrays, i.e., $\tau_{m,n,r} \approx \tau_r$ \cite{1597550}. The delay-DoA representation of the channel is obtained by taking the Fourier transform along the spatial axes as 
\begin{align}
    h_{m,n}(\tau,z,\tilde{z}) = \sum_{r=1}^{R} \alpha_{r} \delta(\tau-\tau_{r}) e^{-\mathrm{j}2\pi (z\theta_{n,r}+\tilde{z}\phi_{m,r})}.\label{eq:radar_channel}
\end{align}
The signal from each transmit antenna is reflected back by these scatterers toward a multi-antenna radar receiver. We discretize the spatial dimension in terms of ULA spacing such that the channel corresponding to the $n$-th transmit antenna at the location $z_n=nd_T$ and $m$-th receive antenna at the location $\tilde{z}_m= {md_R}$ is \cite{van2004detection}
\begin{align} \label{CCIR_1}
    h_{m,n}(\tau) &= \sum_{r=1}^{R} \alpha_{r} \delta(\tau-\tau_{r}) e^{-\mathrm{j}\frac{2\pi}{\lambda} (n d_T\theta_{n,r}+md_R \phi_{m,r})}.
\end{align}
For collocated arrays, DoD and DoA remain constant over the arrays \cite{1597550}. Hence, $\theta_{n,r} \approx \theta_r$ and $\phi_{m,r} = \phi_r$. The channel is
\begin{align} \label{CCIR_2}
    h_{m,n}(\tau) &\approx h(\tau,\theta_{r},\phi_{r})\nonumber\\
    &= \sum_{r=1}^{R} \alpha_{r} \delta(\tau-\tau_{r}) e^{-\mathrm{j}\frac{2\pi}{\lambda} (n d_T\theta_{r}+md_R \phi_{r})}.
\end{align}

The received signal is the convolution of $h(\tau,\theta_{r},\phi_{r})$ with the transmit signal. Accounting for only the transmit array response vector and collecting the echoes of all transmit signals, the received signal trail at the $m$-th antenna for the $k$-th pulse is $\tilde{y}_{m,k}= \sum_{r=1}^{R} \alpha_{r} e^{-\mathrm{j}\frac{2\pi md_R \phi_{r}}{\lambda}} \mathbf{c}^T(\theta_r) \mathbf{x}(t-kT-\tau_{r})$. Denote the $M\times 1$ receive steering vector by $\mathbf{b}(\theta) = [b_1(\theta), \cdots, b_M(\theta)]^T = [1, e^{-j\frac{2\pi d_R}{\lambda}\theta}, \cdots, e^{-j\frac{2\pi (M-1)d_R}{\lambda}\theta}]^T  \in \mathbb{C}^{M \times 1}$. Then, the $N\times 1$ received signal after accounting for the receive array response vector is
\begin{align}
\mathbf{y}_{k}&= \sum_{r=1}^{R} \alpha_{r} \mathbf{b}(\phi_r) \mathbf{c}^T(\theta_r) \mathbf{x}(t-kT-\tau_{r})+\mathbf{v}_{k}(t),\nonumber\\
\end{align}
where $\mathbf{v}_{k}(t)$ is the temporally and spatially additive white Gaussian noise. Again, for a collocated array, $\theta_{r}=\phi_{r}$. After filtering out signals corresponding to each transmit signal in the frequency domain at each receive antenna, define $\mathbf{a}(\theta) = f(\mathbf{b}(\theta),  \mathbf{c}(\theta))$, where $f(\cdot)$ is a function of receive and transmit steering vectors. In the discretized DoA domain, wherein DoAs are aligned to a grid that matches the array resolution, Fourier beamforming produces an impulse at the scatterer location because $\mathbf{a}^H\mathbf{a}=1$. \par
Hence, fixing the direction $\theta_r$, the received signal at the $m$-th antenna for $k$-th pulse is the convolution
\begin{align}
\mathbf{y}_{m,k}(t)= \sum_{n=1}^{N} (\mathbf{x}_n * \mathbf{h}_{m,n,k})(t) + \mathbf{v}_{m,k}(t),\; \forall m,\forall k.
\end{align}
Sampling at the Nyquist rate produces $L=BT$ samples as
\begin{align}
    \mathbf{y}_{m,k}[\tilde{n}]=\sum_{n=1}^{N}\sum_{l} \mathbf{x}_n[l] \mathbf{h}_{m,n,k}[\tilde{n}-l]+\mathbf{v}_{m,k}[\tilde{n}],\; \forall m,\forall k.
\end{align}
Further, assume $\mathbf{x}_n = \left[x_n[0], x_n[1],\hdots,x_n[L-1]\right]^T \in \mathbb{C}^{L \times 1}$ to be the discretized transmit radar signal vector of the $n$-th transmit antenna and $\mathbf{h}_{m,n,k}=\left[{h}_{m,n,k}[0],{h}_{m,n,k}[1],\hdots,{h}_{m,n,k}[R-1]\right]^T\in  \mathbb{C}^{R \times 1}$ as the discretized CCIR vector corresponding to the $m$-th receive, $n$-th transmit and the $k$-th transmit pulse with $R$ range-bins. Then, the corresponding discretized received vector $\mathbf{y}_{m,k} = \left[\mathbf{y}_{m,1,k}, \mathbf{y}_{m,2,k},\hdots,\mathbf{y}_{m,N,k}\right]^T \in  \mathbb{C}^{N(L+R-1) \times 1}$ is \cite{Guerci2016,gogineni2022high}

\begin{align}\label{Eq2_}
\mathbf{y}_{m,k}=&\sum_{n=1}^{N}\mathbf{h}_{m,n,k} \ast \mathbf{x}_n+\mathbf{v}_{m,k},
\end{align}
where the stacked noise vector $\mathbf{v}_{m,k} \in  \mathbb{C}^{N(L+R-1) \times 1}$. 

Rewrite \eqref{Eq2_} as
\begin{align}
\mathbf{y}_{m,k}=&\sum_{n=1}^{N}\mathbf{X}_n\mathbf{h}_{m,n,k}+\mathbf{v}_{m,k},
\end{align}
where 
\begin{align}
\mathbf{X}_n=\begin{bmatrix}
x_n[1] & 0 & \hdots & 0\\
x_n[2] & x_n[1] & \hdots & 0 \\
\vdots & \vdots & \ddots & \vdots \\
x_n[L] & x_n[L-1] & \hdots & x_n[1] \\
0 & x_n[L] & \hdots & x_n[2] \\
\vdots & \vdots & \ddots & \vdots \\
0 & 0 & \hdots & x_n[L]
\end{bmatrix}	 \in \mathbb{C}^{(L+R-1) \times R},
\end{align}
denotes the transmit signal matrix, whose structure is determined by the LFM signal in \eqref{eq:LFM}.  Furthermore, stacking the received vectors $\mathbf{y}_{m,k}$ for all the $1 \leq m \leq M$ receive antennas, one obtains
\begin{align}
    \mathbf{y}_{k}&= \underbrace{\left(\mathbf{I}_M \otimes \mathbf{X} \right)}_{\widetilde{\mathbf{X}}\in \mathbb{C}^{M(L+R-1) \times NMR}}\begin{bmatrix}
        \mathbf{h}_{1,k} \\ \mathbf{h}_{2,k} \\ \vdots \\\mathbf{h}_{M,k}
    \end{bmatrix}+\begin{bmatrix}
        \mathbf{v}_{1,k} \\ \mathbf{v}_{2,k} \\ \vdots \\\mathbf{v}_{M,k}
    \end{bmatrix}, \nonumber\\
    \mathbf{y}_{k}&=\widetilde{\mathbf{X}}\mathbf{h}_{k}+\mathbf{v}_{k},
\end{align}
where $ \mathbf{y}_{k}=\left[ \mathbf{y}^T_{1,k}, \mathbf{y}^T_{2,k}, \hdots, \mathbf{y}^T_{M,k}\right]^T  \in \mathbb{C}^{M(L+R-1) \times 1}$, $\mathbf{X}=\left[\mathbf{X}_1,\mathbf{X}_2,\hdots,\mathbf{X}_N\right] \in \mathbb{C}^{L+R-1 \times NR}$, $\mathbf{h}_{m,k}=\left[\mathbf{h}_{m,1,k}^T,\mathbf{h}_{m,2,k}^T,\hdots,\mathbf{h}_{m,N,k}^T\right]^T \in \mathbb{C}^{NR \times 1}$, $\mathbf{h}_{k}=\left[\mathbf{h}_{1,k}^T,\mathbf{h}_{2,k}^T,\hdots,\mathbf{h}_{M,k}^T\right]^T \in \mathbb{C}^{NMR \times 1}$, and $\mathbf{v}_{k}=\left[\mathbf{v}_{1,k}^T,\mathbf{v}_{2,k}^T,\hdots,\mathbf{v}_{M,k}^T\right]^T \in \mathbb{C}^{M(L+R-1) \times 1}$. Furthermore, for the MMV scenario, we concatenate the received vector $\mathbf{y}_{k}$ for $K$ subsequent pulses to obtain
\begin{align}   \label{Y_MMV}\mathbf{Y}=\left[\mathbf{y}_{1},\mathbf{y}_{2},\hdots,\mathbf{y}_{K}
\right]=\widetilde{\mathbf{X}} \mathbf{H}+\mathbf{V},
\end{align}
where 
\begin{align}
\mathbf{H}&=\left[\mathbf{h}_{1},\mathbf{h}_{2},\hdots,\mathbf{h}_{K}
\right] \in \mathbb{C}^{NMR \times K}, \\
\mathbf{V}&=\left[\mathbf{v}_{1},\mathbf{v}_{2},\hdots,\mathbf{v}_{K}
\right] \in \mathbb{C}^{M(L+R-1) \times K}.
\end{align} 
Since very few range bins are active or present in the entire range bin domain, each CCIR vector $\mathbf{h}_{k}$ is sparse in nature. Further, the sparsity profile of each $\mathbf{h}_k$ is the same, i.e., if a clutter range bin is present for the 1st pulse then it is present for all the pulses, which in turn results in the channel matrix $\mathbf{H}$ being simultaneous or row sparse in nature as shown in Fig. 1. More specifically, if the $i$-th row denoted as $\mathbf{H}_{i.}$ is non-zero, it implies that all the elements of the $i$-th row are non-zero, and vice versa. The number of pulses over which this is valid depends on a number of factors including the motion of the radar platform. Finally, our goal is to estimate the simultaneously sparse CCIR matrix $\mathbf{H}$ from the multiple measurement matrix $\mathbf{Y}$. Toward this end, we employ a Bayesian learning-based framework which is described next.
\begin{figure}
    \centering  \includegraphics[width=0.9\linewidth]{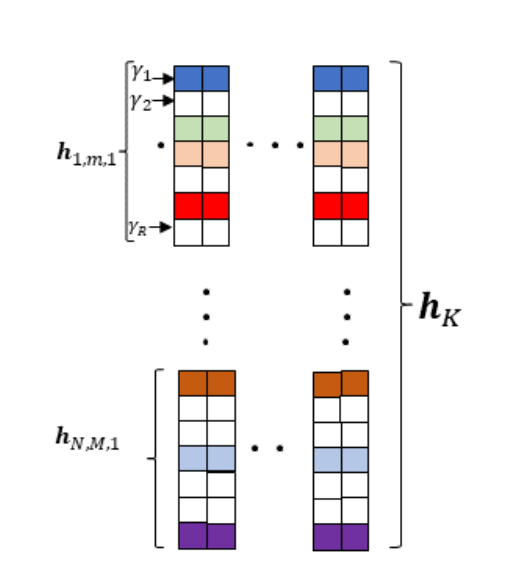}
    \caption{An illustration of simultaneously (row) sparse CCIR matrix.} 
    \label{fig:MSBL}
\end{figure}
\section{SBL-Based Clutter CIR Estimation}
Employing the SBL framework \cite{1315936} on the CCIR matrix is depicted in Fig. \ref{fig:MSBL}, we initially assign the parameterized Gaussian prior to the $i$-th row of the unknown CCIR matrix $\mathbf{H}$ as
\begin{align}
p\left(\mathbf{H}_{i.},{\psi}_i\right)=\prod_{k=1}^{K}\left(\pi \psi_i\right)^{-1} \mathrm{exp}\left(-\frac{|\mathbf{H}_{i.}(k)|^2}{\psi_i}\right),
\end{align}
where the hyperparameter $\psi_i$, $1 \leq i \leq NMR$ denotes the prior variance corresponding to the $i$-th row of $\mathbf{H}$ {which is unknown,} and $\mathbf{\Psi}=\mathrm{diag}\left[\psi_1,\psi_2,\hdots,\psi_{NMR}\right] \in \mathbb{R}^{NMR \times NMR}$ is the diagonal matrix of the hyperparameters.  The overall PDF of the CCIR matrix is 
\begin{align} \label{Prior_H1}p\left(\mathbf{H},\mathbf{\Psi}\right)=\prod_{i=1}^{NMR}\prod_{k=1}^{K}\left(\pi \psi_i\right)^{-1} \mathrm{exp}\left(-\frac{|\mathbf{H}_{i.}(k)|^2}{\psi_i}\right).
\end{align}

Next, one needs to estimate the unknown hyperparameter vector ${\boldsymbol{\psi}}$, which can be estimated by maximizing $\log p(\mathbf{Y};\boldsymbol{\Psi})$ with respect to ${\boldsymbol{\psi}}$, where 
\begin{align}
  \log p(\mathbf{Y};\boldsymbol{\Psi}) =\frac{-K}{2} \log |\mathbf{\Sigma}_y|-\frac{1}{2}\sum_{k=1}^K \mathbf{y}_k^H \boldsymbol{\Sigma}^{-1}_{y}{\mathbf{y}_k}.
\end{align}
 The covariance matrix of the received vector $\mathbf{y}_k$ is $\boldsymbol{\Sigma}_{\mathbf{y}}=\left(\widetilde{\mathbf{X}} \boldsymbol{\Psi} \widetilde{\mathbf{X}}^H + \mathbf{R}_v \right) \in \mathbb{C}^{\tilde{M} \times \tilde{M}}$. From the above equation, it is evident that direct maximization is not possible due to the non-convexity of the objective function. Hence, one needs to rely on the type-II evidence maximization framework which uses expectation-maximization (EM) to obtain the maximum likelihood estimate (MLE) of the hyperparameters. From the prior assignment above, as the hyperparameter $\psi_i \rightarrow 0$, the corresponding row $\mathbf{H}_{i} \rightarrow 0$. Consequently, the estimation of CCIR matrix $\mathbf{H}$ is transferred to the estimation of a hyperparameter vector $\boldsymbol{\psi}$ with the correct number and location of nonzero elements. Hence, the actual sparse matrix $\mathbf{H}$ acts as a latent or hidden variable and is estimated using the available observations $\mathbf{Y}$ and the hyperparameters estimated from the SBL procedure. The next subsection describes the EM algorithm used for this purpose.
\subsection{EM Procedure}
In the $l$-th EM iteration, let $\widehat{\mathbf{\Psi}}^{(l)}$ denote the estimate of the unknown hyperparameter matrix $\mathbf{\Psi}$. \newline
\textbf{E-Step:} In this step, the average values of the latent or hidden variable are calculated given the observation matrix $\mathbf{Y}$ and the current hyperparameter estimate matrix $\widehat{\mathbf{\Psi}}^{(l)}$. In other words, this step helps in filling the missing data or estimating the latent variable. Toward this end, in the $l$-th EM iteration, the average log-likelihood $\mathcal{L}\left(\mathbf{\Psi}|\mathbf{\Psi}^{(l)}\right)$ of the complete data set $\{{\mathbf{Y}},\mathbf{H}\}$ can be evaluated as
\begin{align}\label{LLH}
&\mathcal{L}\left(\mathbf{\Psi}|\widehat{\mathbf{\Psi}}^{(l)}\right)=\mathbb{E}_{\mathbf{H}|{\mathbf{Y}};\widehat{\mathbf{\Psi}}^{(l)}}\left\{\log p\left({\mathbf{Y}},\mathbf{H};\widehat{\mathbf{\Psi}}^{(l)}\right)\right\} \nonumber \\
&=\mathbb{E}_{\mathbf{H}|{\mathbf{Y}};\widehat{\mathbf{\Psi}}^{(l)}}\left\{\log p\left({\mathbf{Y}}|\mathbf{H}\right)+\log p \left(\mathbf{H};\widehat{\mathbf{\Psi}}^{(l)}\right)\right\}. 
\end{align}
{The first term above is
\begin{align}
    &\log p\left({\mathbf{Y}}|\mathbf{H}\right) \nonumber \\&= -\kappa_1 - \sum_{k=1}^{K}\left(\mathbf{Y}_{.k}-\widetilde{\mathbf{X}}\mathbf{H}_{.k}\right)^H\mathbf{R}_v^{-1}\left(\mathbf{Y}_{.k}-\widetilde{\mathbf{X}}\mathbf{H}_{.k}\right),
\end{align}
where $\kappa_1=\log(\pi^{(N+R-1)}|\mathbf{R}_v|)$. The term above is independent of the hyperparameter covariance matrix $\mathbf{\Psi}^{(l)}$ and hence can be ignored in the subsequent M-Step. Furthermore, the second term in \eqref{LLH} reduces to  
\begin{align}\label{Decop}
&\mathbb{E}_{\mathbf{H}|{\mathbf{Y}};\widehat{\mathbf{\Psi}}^{(l)}}\left\{\log p \left(\mathbf{H};\widehat{\mathbf{\Psi}}^{(l)}\right)\right\}\nonumber\\ &=\sum_{i=1}^{NMR}\sum_{k=1}^{K} -\log \left(\pi \psi_i \right)  -\frac{1}{\psi_i}\mathbb{E}_{\mathbf{H}|{\mathbf{Y}};\widehat{\mathbf{\Psi}}^{(l)}}\left\{|\mathbf{H}_{i.}(k)|^2\right\}.
\end{align} 
 \\
\textbf{M-Step:} In this step, the expected log-likelihood function calculated in the E-step is maximized with respect to the hyperparameters. Toward this, the objective function of the M-step can be given as
\begin{align}
\widehat{\boldsymbol{\psi}}^{(l+1)}=\argmax_{\boldsymbol{\psi}} \mathbb{E}_{\mathbf{H}|{\mathbf{Y}};\widehat{\mathbf{\Psi}}^{(l)}}\left\{\log p \left(\mathbf{H};\widehat{\mathbf{\Psi}}^{(l)}\right)\right\}.
\end{align}
In order to evaluate the above conditional expectation, one needs to evaluate the posterior PDF of $\mathbf{H}$ which is given as}
$p\left(\mathbf{H}|{\mathbf{Y}};\widehat{\mathbf{\Psi}}^{(l)}\right) \sim \mathcal{CN}\left(\mathcal{M}^{(l)}, \mathbf{\Sigma}^{(l)}\right)$, where \cite{10.5555/151045}
\begin{align}
\boldsymbol{\mathcal{M}}^{(l)}&= \mathbf{\Sigma}^{(l)}{\widetilde{\mathbf{X}}}^H \mathbf{R}_v^{-1} {\mathbf{Y}} \label{Mean}\\
\mathbf{\Sigma}^{(l)}&=\left(\widetilde{\mathbf{X}}^H\mathbf{R}_v^{-1} \widetilde{\mathbf{X}}+
\left(\widehat{\mathbf{\Psi}}^{(l)}\right)^{-1}\right)^{-1}, \label{Covariance} 
\end{align}
It follows from \eqref{Decop} that the maximization problem is decoupled from each $\psi_i$. It may thus be solved to produce estimates $\widehat{\boldsymbol{\psi}}^{(l+1)}$ in the $l$-th EM iteration as
\begin{align}
\widehat{\psi}_i^{(l+1)}&=\frac{1}{K}\sum_{k=1}^{K} \mathbb{E}_{\mathbf{H}|{\mathbf{Y}};\widehat{\mathbf{\Psi}}^{(l)}}\left\{|\mathbf{H}_{i.}(k)|^2\right\}\nonumber \\&= \mathbf{\Sigma}^{(l)}(i,i)+ \frac{1}{K}\sum_{k=1}^{K}\left|\boldsymbol{\mathcal{M}}^{(l)}_{i.}(k)\right|^2, \forall \; i. \label{Hyp_Est}
\end{align}
{This iterative procedure needs to be repeated till either convergence is achieved or the maximum number of EM iterations denoted by $l_{\max}$ have reached.} Upon convergence, the SBL-based estimate of the sparse CCIR channel $\mathbf{H}$ is given by $\widehat{\mathbf{H}}=\boldsymbol{\mathcal{M}}^{(l)}$. This is the standard SBL procedure to estimate the simultaneously sparse channel matrix $\mathbf{H}$ for the considered MMV scenario \cite{4244754}. However, this leads to a very high computational complexity of the order of $\mathcal{O}\left(N^3M^3R^3\right)$ which is a result of a need to invert a $NMR$ dimensional matrix in the E-step. In this regard, a low complexity version of the SBL algorithm which was recently developed in \cite{9807393} for the single measurement vector scenario known as covariance-free EM (CoFEM) can be invoked. In this work, we extend the CoFEM algorithm for the MMV scenario. In addition, we also take more advanced and practical scenarios of group, joint, and joint-cum-group sparse CCIR matrix estimation for the problem under consideration, which is described next.
\section{Covariance-Free SBL}
One can accelerate EM procedure by avoiding the inversion of the covariance matrix given in \eqref{Covariance}. Also, by using an alternate way to calculate the diagonal elements of $\boldsymbol{\Sigma}$ which are required in the hyperparameter estimation \eqref{Hyp_Est}. This was achieved in \cite{9807393} for the SMV scenario using some linear algebra and is described next for the MMV scenario as opposed to the SMV scenario in \cite{9807393}. The posterior mean matrix equation in \eqref{Mean} can also be equivalently written as
    \begin{align}     \left(\mathbf{\Sigma}^{(l)}\right)^{-1}\mathbf{\mathcal{M}}^{(l)}=\widetilde{\mathbf{X}}\mathbf{R}_v^{-1}\mathbf{Y}.
    \end{align}
    Thus $k$-th column denoted by $\mathbf{\mathcal{M}}^{(l)}_{.k}$ can be seen as the solution of linear system of equations $\mathbf{C}\mathbf{d}=\mathbf{z}$, where $\mathbf{C}=\left(\mathbf{\Sigma}^{(l)}\right)^{-1}$, $\mathbf{d}=\mathbf{\mathcal{M}}^{(l)}_{.k}$ and $\mathbf{z}=\widetilde{\mathbf{X}}\mathbf{R}_v^{-1}\mathbf{Y}_{.k}$. It can be solved by the system of linear equations using conjugate gradient (CG)-algorithm \cite{hestenes1952methods}. Furthermore, one can find the diagonal elements of $\mathbf{\Sigma}^{(l)}$ employing the procedure procedure proposed in \cite{bekas2007estimator}, which are required in M-step. We summarize the diagonal estimation procedure in the next subsection.
   \subsection{Diagonal Estimation Rule}
{For a square matrix $\mathbf{S} \in \mathbb{C}^{q \times q}$, an unbiased estimator of the diagonal elements, which is denoted by $\mathbf{e}$ can be evaluated using
\begin{align}
    \mathbf{e}=\left(\sum_{\tilde{l}=1}^{\tilde{L}}\mathbf{u}_{\tilde{l}} \odot \mathbf{w}_{\tilde{l}}\right) \oslash \left(\sum_{\tilde{l}=1}^{\tilde{L}}\mathbf{u}_{\tilde{l}} \odot \mathbf{u}_{\tilde{l}}\right),
\end{align}
where $\mathbf{w}_{\tilde{l}}=\mathbf{S}\mathbf{u}_{\tilde{l}}$, while $\mathbf{u}_1,\mathbf{u}_2,\hdots,\mathbf{u}_{\tilde{L}} \in \mathbb{R}^{q \times 1}$ denote ${\tilde{L}}$ probing vectors, consisting of independent and identically distributed (i.i.d.) elements, and satisfying  $\mathbb{E}[\mathbf{u}_{j}(a)]=0$ for $a=1,\ldots,q$. The probing vectors $\mathbf{u}_{\tilde{l}}$ for $l=1,2,\hdots,{\tilde{L}}$ can be generated using Rademacher distribution which in turn implies that each element of the $\mathbf{u}_{\tilde{l}}$ will be either $+1$ or $-1$ with equal probability. Then, each element $\mathbf{e}(j)$ is an unbiased estimator of $\mathbf{S}(j,j)$ and  $\mathbf{E}\left[\mathbf{e}(j)\right]=\mathbf{S}(j,j)$.
    
        Hence, one can write the diagonal estimation rule in a compact form as
        \begin{align} \label{Comp_e}
        \mathbf{e} = \frac{1}{{\tilde{L}}} \sum _{i=1}^{{\tilde{L}}} \mathbf{u}_{\tilde{l}} \odot \mathbf{w}_{\tilde{l}}. 
        \end{align} 
        In order to utilize this rule to estimate the diagonal elements of $\mathbf{\Sigma}^{(l)}$, one needs to multiply $\mathbf{\Sigma}^{(l)}$ to each vector $\mathbf{u}_{\tilde{l}}$ to obtain $\mathbf{w}_{\tilde{l}}$; this can be done once again by solving a linear system $\left(\mathbf{\Sigma}^{(l)}\right)^{-1}\mathbf{w}_{\tilde{l}}=\mathbf{z}_{\tilde{l}}$, where $\mathbf{z}_{\tilde{l}}:=\mathbf{u}_{\tilde{l}}$. Upon solving these $L+K$ parallel linear system of equations one can obtain the matrix ${\mathbf{\mathcal{M}}}$ and diagonal elements $\mathbf{\Sigma}(j,j)$, for all $j$ with lower computational complexity to simplify the computationally heavy E-step. Another interesting observation is that one can obtain the posterior mean matrix $\bf{\mathcal{M}}^{(l)}$ and $\mathbf{w}_{\tilde{l}}$ by employing CG-algorithm to solve the linear system of equations given as $\mathbf{C}\mathbf{W}=\mathbf{Z}$, where matrices $\mathbf{C}\in \mathbb{C}^{(L+R-1)M \times K}$ and $\mathbf{Z}\in \mathbb{C}^{(L+R-1)M\times ({\tilde{L}}+K)}$ are defined as follows:
      \begin{align} 
      {\mathbf{C}} &:= \left(\mathbf{\Sigma}^{(l)}\right)^{-1}=\left(\widetilde{\mathbf{X}}^H \mathbf{R}_v^{-1}\widetilde{\mathbf{X}}+ \left(\mathbf{\Psi}^{(l)}\right)^{-1} \right), \label{Eq:A}\\ {\mathbf{Z}} &:= \begin{bmatrix}\boldsymbol{u}_{1} \:\vert \:\boldsymbol{u}_{2} \:\vert \:\ldots \:\vert \:\boldsymbol{u}_{{\tilde{L}}} \:\vert \:\widetilde{\mathbf{X}}^H \mathbf{R}_v^{-1} \boldsymbol{Y} \label{Eq:B} \end{bmatrix}.  
      \end{align}               
    Once, we obtain the knowledge of $\boldsymbol{\mathcal{M}}^{(l)}$ and diagonal elements of $\mathbf{\Sigma}^{(l)}$, one can implement Eq.\eqref{Hyp_Est} to obtain the estimates of the hyperparameter vector $\boldsymbol{\psi}$. It has only a complexity of $\mathcal{O}\left(V {\tilde{L}}\tau\right)$, where $V$ represents the number CG-steps required to solve linear system of equations, ${\tilde{L}}$ represents the number of probe vectors and $\tau$ represents the time required to calculate $\sigma_v^{-2}\widetilde{\mathbf{X}}^H \widetilde{\mathbf{X}}$ which is $\mathcal{O}\left(NMR \log(NMR)\right)$.} Algorithm 1 summarizes the proposed CoFBL simultaneously sparse CCIR matrix estimation algorithm for the MMV scenario. 
\begin{algorithm}[H]
	\caption{Covariance-free SBL-based sparse CCIR estimation} 
	\label{alg:sbl}
	\begin{algorithmic}[1]
		\Statex \textbf{Input:} ${\mathbf{Y}}, \widetilde{\mathbf{X}}$, $\epsilon$ (Threshold),  $l_{\text{max}}$ 
		\Statex \textbf{Output:} $\widehat{\mathbf{H}}$
        \State set $l=1$ and $\widehat{\mathbf{\Psi}}^{(0)}=\mathbf{I}$
        
        \State \textbf{E-step:} Compute $\mathbf{C}$ using \eqref{Eq:A} 
        \State Draw $\mathbf{u}_1$, $\mathbf{u}_2$,...,$\mathbf{u}_{\tilde{L}}$ using Rademacher distribution
        \State Compute ${\mathbf{Z}}$ using \eqref{Eq:B}
        \State $\left[\mathbf{w}_1|\mathbf{w}_2|\hdots|\mathbf{w}_{{\tilde{L}}}|\boldsymbol{\mathcal{M}}\right]=\mathrm{CG}(\bf{C},\bf{Z})$
        \State Compute $\mathbf{e}$ using \eqref{Comp_e}
        \State \textbf{M-Step:} Compute $\widehat{\psi}_i^{(l+1)}$ using \eqref{Hyp_Est}, $\forall \; i$ 
        \State if $||\hat{\boldsymbol{\psi}}^{(l+1)}-\hat{\boldsymbol{\psi}}^{(l)}|| \geq \epsilon$ or $l \leq l_{\text{max}}$ 
        \State Exit
        \State end
        \State \textbf{return} $\widehat{\mathbf{H}}=\boldsymbol{\mathcal{M}}^{(l+1)}$.
 \end{algorithmic}
\end{algorithm}
{If a particular range bin continues to be populated over a certain number of pulses (stationarity) and across transmit-receive antenna pairs (co-located). Further, the channel vector for the $k$-th pulse, $\mathbf{h}_k \in \mathbb{C}^{NMR \times 1}$, considered in this paper is obtained by stacking across the transmit antennas and followed by the receive antennas. Thus, coupled with the aforementioned fact renders structure to the sparsity, which we have attempted to exploit in Section-V.} We describe three important scenarios where the simultaneously sparse CCIR matrix exhibits group, joint, and joint-cum-group sparse structure, respectively, which can be exploited further to reduce the number of hyperparameter to estimate significantly. 
\section{Diverse Sparsity Scenarios}
The clutter patch comprising of multiple range bins and its echoes are collected over multiple pulses. This gives rise to a variety of sparse clutter scenarios encountered in practice. In the sequel, we discuss these models.
\subsection{Group Sparsity}
\begin{figure*}[t]
    \centering  \includegraphics[width=\linewidth]{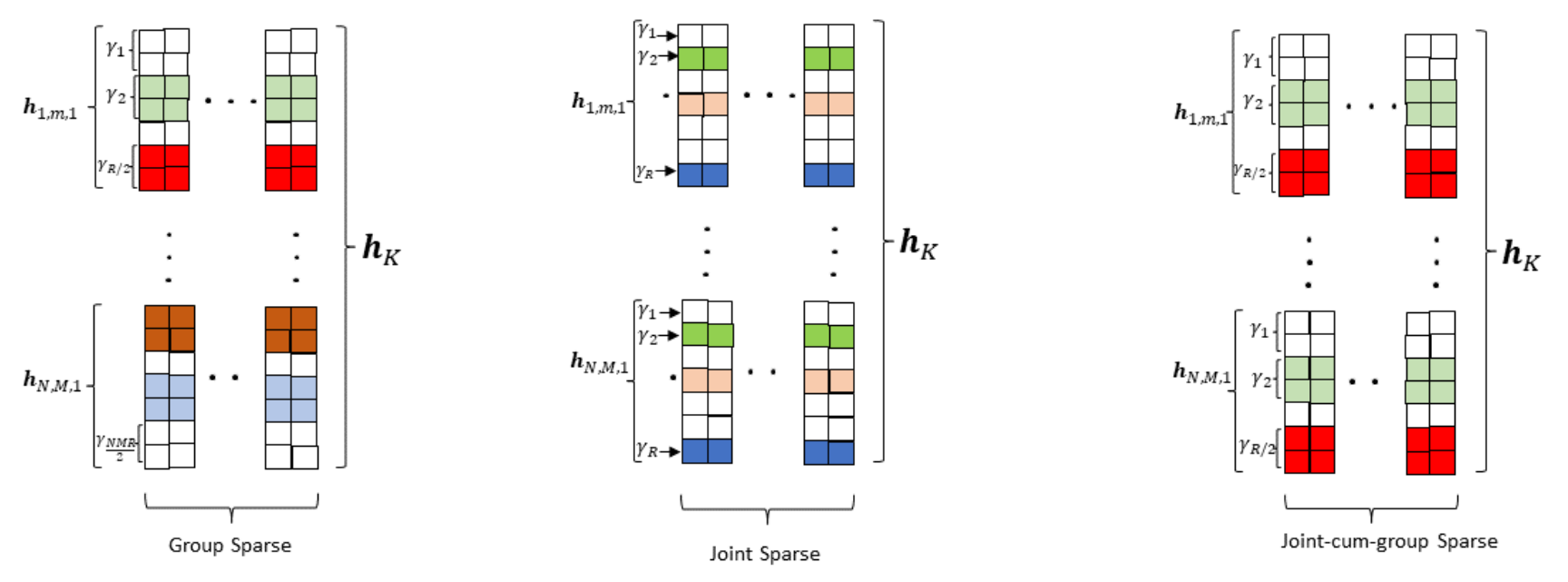}
    \caption{Illustrations of group, joint, and joint-cum-group sparse CCIR matrices.}
    \label{fig:JGMSBL}
\end{figure*}

The parameterized Gaussian prior of the $i$-th group of the unknown CCIR matrix $\mathbf{H}$ as
\begin{align}
p\left(\mathbf{H}_{i.},\bar{\psi}_i\right)=\prod_{k=1}^{K}\left(\pi \bar{\psi}_i\right)^{-d} \mathrm{exp}\left(-\frac{\sum_{(i-1)d+1}^{id}|\mathbf{H}_{i.}(k)|^2}{\bar{\psi}_i}\right),
\end{align}
where $d$ is the group sparsity cluster length, the hyperparameter $\bar{\psi}_i$, $1 \leq i \leq NMR/d$ denotes the prior variance corresponding to the $i$-th group of $\mathbf{H}$ {which is unknown}, $i$-th group contains indices from $(i-1)d+1$-th row to $id$-th row, and $\bar{\mathbf{\Psi}}=\mathrm{diag}\left[\bar{\psi}_1,\bar{\psi}_2,\hdots,\bar{\psi}_{NMR/d}\right] \in \mathbb{R}^{NMR/d \times NMR/d}$ is the diagonal matrix of the hyperparameters. The CCIR matrix is depicted in Fig. \ref{fig:JGMSBL}. The overall PDF of the CCIR matrix is 
\begin{align} \label{Prior_H2}
&p\left(\mathbf{H},{\mathbf{\Psi}}_G\right) \nonumber \\&=\prod_{i=1}^{NMR/d}\prod_{k=1}^{K}\left(\pi \bar{\psi}_i\right)^{-d} \mathrm{exp}\left(-\frac{\sum_{(i-1)d+1}^{id}|\mathbf{H}_{i.}(k)|^2}{\bar{\psi}_i}\right),
\end{align}
The hyperparameter covariance matrix ${\boldsymbol{\Psi}}_G$ for this scenario is given as
\begin{align}
    {\boldsymbol{\Psi}}_G= \bar{\mathbf{\Psi}} \otimes \mathbf{I}_d \in \mathbb{C}^{NMR \times NMR}
\end{align}
Theorem 1 summarizes the hyperparameter estimate for this scenario.
\begin{theo}
In the $l$-th EM iteration, given  $\widehat{\bar{\psi}}_i^{(l-1)}$, $1 \leq i \leq NMR/d$, the current hyperparameter estimates $\widehat{\bar{\psi}}_i^{(l)}$ which maximizes $\mathbb{E}_{\mathbf{H}|{\mathbf{Y}};\widehat{\mathbf{\Psi}}^{(l)}_G}\left\{\log p\left({\mathbf{Y}},\mathbf{H};\widehat{\mathbf{\Psi}}^{(l)}_G\right)\right\}$ can be evaluated as
\begin{align} \label{Para_Est_GMSBL}
&\widehat{\bar{\psi}}_i^{(l)}=\frac {1}{dK} \sum _{k=1}^{K}|\mathcal{M}^{(l)}_G\left(\tilde{m},k\right)|^2 +\frac {1}{d} \boldsymbol{\Sigma}^{(l)}_G\left(\tilde{m},\tilde{m}\right),
\end{align}
where $\tilde{m}=(i-1)d:id$, $\boldsymbol{\mathcal{M}}^{(l)}_G= \left(\mathbf{\Sigma}^{(l)}_G\right)^{-1} {\widetilde{\mathbf{X}}}^H \mathbf{R}_v^{-1} {\mathbf{Y}} \in \mathbb{C}^{NMR \times K}$, $\mathbf{\Sigma}^{(l)}_G=\left(\widetilde{\mathbf{X}}\mathbf{R}_v^{-1} \widetilde{\mathbf{X}}^H+
\left(\widehat{\mathbf{\Psi}}^{(l)}_G\right)^{-1}\right)^{-1} \in \mathbb{C}^{NMR \times NMR}$ are the posterior mean and covariance matrices, respectively. 
\end{theo}
\begin{IEEEproof}
The detailed derivation is given in Appendix A. 
\end{IEEEproof}
This algorithm reduces the total number of hyperparameters to be estimated by a factor of $d$ as compared to the CoFBL algorithm. The subsequent subsection examines the situation in which the CCIR matrix has a joint sparse structure, where the channel vector $\mathbf{h}_{n,m,k}$ exhibits the same sparsity structure for each $n$, $m$ and $k$. 
\subsection{Joint Sparsity}
As depicted in Fig. \ref{fig:JGMSBL}, assuming the hyperparameter $\tilde{\psi}_i$ represents the sparsity of the $i$-th range bin of each channel vector $\mathbf{h}_{n,m,k}$ for all $n$, $m$ and $k$. Hence, the prior PDF corresponding to each channel vector $\mathbf{h}_{n,m,k}$ is
\begin{align}
    p\left(\mathbf{h}_{n,m,k},\tilde{\boldsymbol{\psi}}\right)=\prod_{i=1}^{R}\left(\pi \tilde{\psi}_i\right)^{-1} \mathrm{exp}\left(-\frac{|\mathbf{h}_{n,m,k}(i)|^2}{\tilde{\psi}_i}\right),  
\end{align}
where $\tilde{\boldsymbol{\psi}}=[\tilde{\psi}_1,\tilde{\psi}_2,\hdots,\tilde{\psi}_R]^T \in \mathbb{R}^{R \times 1}$ and $\widetilde{\mathbf{\Psi}}=\textrm{diag}\left[\tilde{\boldsymbol{\psi}}\right] \in \mathbb{R}^{R \times R}$. Let the hyperparameter covariance matrix for this scenario is 
\begin{align}   {\mathbf{\Psi}}_J=\left(\mathbf{I}_M \otimes \left( \mathbf{I}_N
\otimes \widetilde{\mathbf{\Psi}} \right) \right) \in \mathbb{C}^{NMR \times NMR}
\end{align}
Hence, the prior PDF of the channel matrix $\mathbf{H}$ is\par\noindent\small
\begin{align}\label{Prior_JSBL}
    p\left(\mathbf{H},\mathbf{\Psi}_J\right)=\prod_{n=1}^{N}\prod_{m=1}^{M}\prod_{k=1}^{K}\prod_{i=1}^{R}\left(\pi \tilde{\psi}_i\right)^{-1} \mathrm{exp}\left(-\frac{|\mathbf{h}_{n,m,k}(i)|^2}{\tilde{\psi}_i}\right),
\end{align}\normalsize
The following theorem summarizes the hyperparameter estimates for this scenario. 
\begin{theo}
In the $l$-th EM iteration, given $\widehat{\tilde{\psi}}_i^{(l-1)}$, $1 \leq i \leq R$, the current hyperparameter estimates $\widehat{\tilde{\psi}}_i^{(l)}$ which maximizes $\mathbb{E}_{\mathbf{H}|{\mathbf{Y}};\widehat{\mathbf{\Psi}}^{(l)}_J}\left\{\log p\left({\mathbf{Y}},\mathbf{H};\widehat{\mathbf{\Psi}}^{(l)}_J\right)\right\}$ can be evaluated as
\begin{align} \label{Para_Est_JMSBL}
&\widehat{\tilde{\psi}}_i^{(l)}=\frac {1}{NMK} \sum _{n=1}^{N} \sum _{m=1}^{M} \sum _{k=1}^{K}|\bar{\mathcal{M}}^{(l)}_J\left(\bar{m}+i,k\right)|^2 \nonumber \\&+\frac {1}{NM} \sum _{n=1}^{N} \sum _{m=1}^{M}\bar{\boldsymbol{\Sigma}}^{(l)}_J\left(\bar{m}+i,\bar{m}+i\right),
\end{align}
where $\bar{m}=(m-1)NR+(n-1)R$.
\end{theo}
\begin{IEEEproof}
See Appendix B.
\end{IEEEproof}

The quantities $\bar{\mathcal{M}}^{(l)}_J$ and $\bar{\boldsymbol{\Sigma}}^{(l)}_J$ represent the posterior mean and covariance matrices for the CCIR matrix $\mathbf{H}$ for the joint sparse scenario, which can be obtained from (26) and (27), respectively by replacing the hyperparameter covariance matrix $\widehat{\mathbf{\Psi}}$ by $\widehat{\boldsymbol{\Psi}}_J$. The detailed proof of the above theorem is given in Appendix B. Please note that in each EM iteration, the total number of hyperparameters that one needs to estimate reduces to $R$ in contrast to the $NMR$ for the CoFBL algorithm. This results in significant computational resource savings and makes the algorithm faster.
\subsection{Joint-Cum-Group Sparsity}
Furthermore, the clutter patches generally occur in patches and hence if a particular range bin is active then there is a very high possibility that the nearby range bins are also active in practice. This makes the unknown sparse CCIR matrix joint-cum-group sparse in nature as depicted in Fig. \ref{fig:JGMSBL}. Therefore, the parameterized Gaussian prior to the clutter channel $\mathbf{h}_{n,m,k}$ for this scenario can be assigned as
\begin{equation*} 
p(\mathbf {h}_{n,m,k}; \underline{\boldsymbol{\psi}}) = \prod _{i=1}^{R/d}\frac {1}{(\pi \underline{\psi}_{i})^d} \exp \left({- \displaystyle \frac {\sum_{(i-1)d+1}^{id}\vert \mathbf {h}_{n,m,k}(i)\vert ^{2}}{\underline{\psi} _{i}}}\right),
\end{equation*}
\begin{algorithm}
	\caption{Covariance-free SBL-based joint, group, and joint-cum-group sparse CCIR estimation} 
	\label{alg:joint}
	\begin{algorithmic}[1]
		\Statex \textbf{Input:} ${\mathbf{Y}}, \widetilde{\mathbf{X}}$, $\epsilon$ (Threshold),  $l_{\text{max}}$ 
		\Statex \textbf{Output:} $\widehat{\mathbf{H}}$  
        \Statex \textbf{Group Sparse}
        \State set $l=1$ and $\widehat{\mathbf{\Psi}}_G^{(0)}=\mathbf{I}$
        \State for $||\hat{\bar{\boldsymbol{\psi}}}^{(l+1)}-\hat{\bar{\boldsymbol{\psi}}}^{(l)}|| \geq \epsilon$ or $l \leq l_{\text{max}}$     
        \State \textbf{E-step:} Compute $\mathbf{C}=\left(\widetilde{\mathbf{X}}^H \mathbf{R}_v^{-1}\widetilde{\mathbf{X}}+ \left(\widehat{\mathbf{\Psi}}_G^{(l)}\right)^{-1} \right)^{-1}$ 
        \State Draw $\mathbf{u}_1$, $\mathbf{u}_2$,...,$\mathbf{u}_{\tilde{L}}$ using Rademacher distribution
        \State Compute ${\mathbf{Z}}$ using \eqref{Eq:B}
        \State $\left[\mathbf{w}_1,\mathbf{w}_2,\hdots,\mathbf{w}_L|\mathbf{\mathcal{M}}\right]=\mathrm{CG}(\mathbf{C},\mathbf{Z})$
        \State Compute $\mathbf{e}$ using \eqref{Comp_e}
        \State \textbf{M-Step:} Compute $\widehat{\psi}_i^{(l+1)}$ using \eqref{Para_Est_GMSBL}, $\forall \; i$          
        \State end
         \State \textbf{return} $\widehat{\mathbf{H}}=\boldsymbol{\mathcal{M}}_G^{(l+1)}$.
        \Statex \textbf{Joint Sparse}
        \State set $l=1$ and $\widehat{\mathbf{\Psi}}_J^{(0)}=\mathbf{I}$
        \State for $||\hat{\tilde{\boldsymbol{\psi}}}^{(l+1)}-\hat{\tilde{\boldsymbol{\psi}}}^{(l)}|| \geq \epsilon$ or $l \leq l_{\text{max}}$     
        \State \textbf{E-step:} Compute $\mathbf{C}=\left(\widetilde{\mathbf{X}}^H \mathbf{R}_v^{-1}\widetilde{\mathbf{X}}+ \left(\widehat{\widetilde{\mathbf{\Psi}}}_J^{(l)}\right)^{-1} \right)^{-1}$ 
        \State Draw $\mathbf{u}_1$, $\mathbf{u}_2$,...,$\mathbf{u}_{\tilde{L}}$ using Rademacher distribution
        \State Compute ${\mathbf{Z}}$ using \eqref{Eq:B}
        \State $\left[\mathbf{w}_1,\mathbf{w}_2,\hdots,\mathbf{w}_L|\mathbf{\mathcal{M}}\right]=\mathrm{CG}(\mathbf{C},\mathbf{Z})$
        \State Compute $\mathbf{e}$ using \eqref{Comp_e}
        \State \textbf{M-Step:} Compute $\widehat{\psi}_i^{(l+1)}$ using \eqref{Para_Est_JMSBL}, $\forall \; i$          
        \State end
         \State \textbf{return} $\widehat{\mathbf{H}}=\boldsymbol{\mathcal{M}}_J^{(l+1)}$.
        \Statex \textbf{Joint-Cum-Group Sparse}
        \State set $l=1$ and $\widehat{\mathbf{\Psi}}_{JG}^{(0)}=\mathbf{I}$
        \State for $||\hat{\boldsymbol{\underline{\psi}}}^{(l+1)}-\hat{\boldsymbol{\underline{\psi}}}^{(l)}|| \geq \epsilon$ or $l \leq l_{\text{max}}$     
        \State \textbf{E-step:} Compute $\mathbf{C}=\left(\widetilde{\mathbf{X}}^H \mathbf{R}_v^{-1}\widetilde{\mathbf{X}}+ \left(\widehat{\mathbf{\Psi}}_{JG}^{(l)}\right)^{-1} \right)^{-1}$ 
        \State Draw $\mathbf{u}_1$, $\mathbf{u}_2$,...,$\mathbf{u}_{\tilde{L}}$ using Rademacher distribution
        \State Compute ${\mathbf{Z}}$ using \eqref{Eq:B}
        \State $\left[\mathbf{w}_1,\mathbf{w}_2,\hdots,\mathbf{w}_L|\mathbf{\mathcal{M}}\right]=\mathrm{CG}(\mathbf{C},\mathbf{Z})$
        \State Compute $\mathbf{e}$ using \eqref{Comp_e}
        \State \textbf{M-Step:} Compute $\widehat{\underline{\psi}}_i^{(l+1)}$ using \eqref{Para_Est_JGMSBL}, $\forall \; i$          
        \State end
        \State \textbf{return} $\widehat{\mathbf{H}}=\boldsymbol{\mathcal{M}}_{JG}^{(l+1)}$.
 \end{algorithmic}
\end{algorithm} 
where $d$ represents the group sparsity cluster length, $\underline{\boldsymbol{\psi}}=[\underline{\psi}_1,\underline{\psi}_2,\hdots,\underline{\psi}_{\frac{R}{d}}]^T \in \mathbb{R}^{\frac{R}{d} \times 1 }$, and $\underline{\boldsymbol{\Psi}}=\mathrm{diag}\left(\underline{\boldsymbol{\psi}}\right)  \in \mathbb{R}^{\frac{R}{d} \times \frac{R}{d}}$. The prior corresponding to the MMV clutter channel matrix $\mathbf{H} \in \mathbb{C}^{NMR \times K}$ for this scenario is given as\par\noindent\small
\begin{align} \label{Prior_JGSBL}
&p \left ({ {\mathbf {H}}; \underline{\boldsymbol{\Psi}}_{JG}}\right) \nonumber \\ &= \prod _{n=1}^{N} \prod _{m=1}^{M} \prod _{k=1}^{K} \prod _{i=1}^{R/d}\frac {1}{(\pi \underline{\psi}_{i})^d} \exp \left({-\displaystyle \frac {\sum_{(i-1)d+1}^{id}\vert \mathbf {h}_{n,m,k}(i)\vert ^{2}}{\underline{\psi}_{i}}}\right). 
\end{align}\normalsize
The hyperparameter covariance matrix $\boldsymbol{\Psi}_{JG}$ for this scenario is given as
\begin{align}   \boldsymbol{\Psi}_{JG}=\left(\mathbf{I}_M \otimes \left( \mathbf{I}_N
\otimes \left(\underline{\boldsymbol{\Psi}} \otimes \mathbf{I}_d \right) \right) \right) \in \mathbb{C}^{NMR \times NMR}
\end{align}
Theorem 3 summarizes the hyperparameter estimate for this scenario.
\begin{theo}
In the $l$-th EM iteration, given $\widehat{\underline{\psi}}_i^{(l-1)}$, $1 \leq i \leq R/n$, the current hyperparameter estimates $\widehat{\underline{\psi}}_i^{(l)}$ which maximizes $\mathbb{E}_{\mathbf{H}|{\mathbf{Y}};\widehat{\mathbf{\Psi}}^{(l)}}\left\{\log p\left({\mathbf{Y}},\mathbf{H};\widehat{\mathbf{\Psi}}^{(l)}_{JG}\right)\right\}$ can be evaluated as
\begin{align} \label{Para_Est_JGMSBL}
&\widehat{\underline{\psi}}_i^{(l)}=\frac {1}{dNMK} \sum _{n=1}^{N} \sum _{m=1}^{M} \sum _{k=1}^{K}|\mathcal{M}^{(l)}_{JG}\left(\underline{m},k\right)|^2 \nonumber \\&+\frac {1}{dNM} \sum _{n=1}^{N} \sum _{m=1}^{M}\boldsymbol{\Sigma}^{(l)}_{JG}\left(\underline{m},\underline{m}\right),
\end{align}
where $\underline{m}=\bar{m}+(i-1)d:\bar{m}+id$.
\end{theo}
\begin{IEEEproof}
The detailed derivation is given in Appendix C. 
\end{IEEEproof}
In contrast to CoFBL, which estimates total $NMR$ hyperparameters in each SBL iteration, and CoFGBL which requires to estimate $\frac{NMR}{d}$ hyperparameter in each EM iteration, and CoFJBL which requires to estimate $R$ hyperparameter in each EM iteration, the proposed CoFJGBL requires only $R/d$ hyperparameters to be estimated. This further reduces computational complexity and enhances the overall efficient resource utilization. Algorithm 2 summarizes the different CoFBL sparse CCIR matrix estimation algorithms for MMV scenarios. {While some apriori information about the scene to be estimated would help in the selection of appropriate technique, one could undertake the following approach. A few SBL iterations using the CoFBL technique can be used initially to estimate CCIR, and based on the emergence of sparsity structure, one could use COFGBL, CoFJBL and COFGJBL techniques to leverage additional degree of freedom.}
The next sub-section briefly discusses the computational complexity of the proposed algorithms.
\subsection{Computational Complexity}
\begin{table}
\centering
\caption{Computational complexities of the various proposed algorithms}\label{Table_comp}
\begin{tabular}{||l|c|c|}
\hline
Algorithm &  Conventional EM-based SBL & Proposed methods \\
\hline
\hline
CoFBL & $\mathcal{O}\left(N^3M^3R^3\right)$ & $\mathcal{O}\left(V_M \tilde{L}_M\tau_M\right)$\\
\hline
CoFGBL   & $\mathcal{O}\left(N^3M^3R^3\right)$ &  $\mathcal{O}\left(V_G\tilde{L}_G\tau_G\right)$\\
\hline
CoFJBL & $\mathcal{O}\left(N^3M^3R^3\right)$ &  $\mathcal{O}\left(V_J \tilde{L}_J\tau_J\right)$  \\
\hline
CoFJGBL & $\mathcal{O}\left(N^3M^3R^3\right)$ &   $\mathcal{O}\left(V_{JG}\tilde{L}_{JG}\tau_{JG}\right)$ \\
\hline
\end{tabular}
\end{table} 
The per EM iteration computational complexities of the various proposed scheme for the conventional SBL-based implementation is given in table \ref{Table_comp}. The computational complexity reduces since the number of hyperparameters that are to be estimated in every algorithm reduces as a result of exploiting the additional structural sparsity. It can be further reduced with the covariance-free EM-based implementation, which leads to the computational complexity of the $\mathcal{O}\left(U_i\tilde{L}_i\tau_i\right)$, where $U_i$ is the number of CGD steps, $\tilde{L}_i$ is the total number of probe vectors and $\tau_i$ is the complexity of the matrix multiplication $\widetilde{\mathbf{X}}\widetilde{\mathbf{X}}^H$ in each of the proposed algorithm, where $i \in \{M,G,J,JG\}$ represents the different algorithms. {Utilizing the interpretable nature and computational efficiency of CoFBL, CoFAR systems achieve accurate clutter CIR estimation while ensuring transparency in the underlying processes, unlike deep learning methods. This supports CoFAR's goal of conserving resources for environment knowledge acquisition while prioritizing radar surveillance tasks.} 

{The primary goal of the CoFBL algorithm is to reduce computational complexity when estimating large-dimensional sparse quantities \cite{9807393}. If the sparse quantity is not sufficiently large, the computational advantage of CoFBL will be minimal. Moreover, the computational complexity of the CoFBL algorithm is influenced by the number of conjugate gradient steps needed to solve the linear system of equations. Therefore, if the condition number of the matrix ${\mathbf{C}} =\left(\widetilde{\mathbf{X}}^H \mathbf{R}_v^{-1}\widetilde{\mathbf{X}}+ \left(\mathbf{\Psi}^{(l)}\right)^{-1} \right)$ is very high, a preconditioner matrix may be necessary to accelerate the convergence of the conjugate gradient algorithm \cite{9807393}. Nevertheless, the number of conjugate gradient iterations $V$ will always be less than or equal to $NMR$, the length of the unknown sparse quantity \cite{hestenes1952methods}.}

\section{Bayesian Cram\'{e}r-Rao Bounds (BCRBs)}
In this section, we derived BCRB bound for the various algorithms proposed in this work. For any unbiased estimation technique, BCRB serves as a lower bound on the MSE. 
\begin{propo}
The BCRB for the CoFBL technique is \par\noindent\small
\begin{align*} 
\mathrm {E}\left[\hspace {-0.1cm}\left \|{ \widehat {\underline {\mathbf {H}}}-\underline {\mathbf {H}} }\right \|_{F}^{2}\right]\geq& \mathrm{Tr}\left[\hspace {-0.1cm}\left ({\sigma ^{-2}\left (\widetilde{\mathbf{X}}^{H}\widetilde{\mathbf{X}}\right)\otimes \mathbf {I}_{K}+\boldsymbol {\Psi }^{-1}\otimes \mathbf {I}_{K}}\right)^{-1}\right]
\end{align*}\normalsize
\end{propo}
\begin{IEEEproof}
Contrary to the CRB framework, which does not take prior PDF of the unknown quantity into account, BL framework \cite{1315936} assigns a parametric prior PDF in \eqref{Prior_H2}, and it serves a lower bound on the estimation performance of the CoFBL algorithm. Employing the BCRB framework, the total Fisher Information Matrix (FIM) $\mathbf {J}_{T}$ for the unknown CCIR matrix $\mathbf{H}$ is defined as $\mathbf {J}_{T}=\mathbf {J}_{D}+\mathbf {J}_{P}$, where $\mathbf {J}_{D}$ and $\mathbf {J}_{P}$ are the FIMs of the observed data and prior, respectively. After vectorizing Eq. \eqref{Y_MMV}, and defining $\bar{\mathbf{X}}=\left(\mathbf{I}_K \otimes \tilde{\mathbf{X}}\right) \in \mathbb{C}^{MK(K+R-1) \times KNMR}$, one obtains $\mathbf{y}=\bar{\mathbf{X}}\mathbf{h}+\bar{\mathbf{v}}$. Hence, $\mathbf {J}_{T}$ can be mathematically represented as
\begin{align} 
\mathbf {J}_{T} = \underbrace {-\mathrm {E}_{\left ({\mathbf {\underline {y}}, \mathbf {\underline {h}}}\right)}\left[ \frac {\partial ^{2}\mathcal {L}\left ({\mathbf {\underline {y}}|\mathbf {\underline {h}}; {\mathbf {P}}}\right)}{\partial \mathbf {\underline {h}}\partial \mathbf {\underline {h}}^{H}}\right]}_{\mathbf {J}_{D}} \underbrace {-\mathrm {E}_{\mathbf {\underline {h}}}\left[ \frac {\partial ^{2}\mathcal {L}\left ({\mathbf {\underline {h}};{\mathbf {P}}}\right)}{\partial \mathbf {\underline {h}}\partial \mathbf {\underline {h}}^{H}}\right]}_{\mathbf {J}_{P}}, 
\end{align}
where $\mathbf{y}=\mathrm{vec}\left(\mathbf{Y}\right) \in \mathbb{C}^{K(K+R-1) \times 1}$ and $\mathbf{h}=\mathrm{vec}\left(\mathbf{H}\right) \in \mathbb{C}^{KNMR \times 1}$, and $\mathbf{P}=\mathbb{E}[\mathbf{h}\mathbf{h}^H] \in \mathbb{C}^{KNMR \times KNMR}$. The log-likelihoods of the observation vector $\mathbf{y}$ and CCIR vector $\mathbf{h}$ are defined as $\mathcal {L}\left({\mathbf {\underline {y}}|\mathbf {\underline {h}};{\mathbf {P}}}\right)=\log p({\mathbf {\underline {y}}|\mathbf {\underline {h}};{\mathbf {P}}})$ and $\mathcal {L}\left ({\mathbf {\underline {h}};{\mathbf {P}}}\right)=\log \left ({\mathbf {\underline {h}};{\mathbf {P}}}\right)$, respectively, which are parameterized by $\mathbf{P}$. Using \eqref{Prior_H1}, the log-likelihood is given as
\begin{equation*} 
\mathcal {L}\left ({\mathbf {\underline {h}};{\mathbf {P}}}\right) = \left ({c-\mathbf {\underline {h}}^{H}{\mathbf {P}}^{-1}\mathbf {\underline {h}}}\right)\!,
\end{equation*}
where $c=-NMK \log(\pi)-\log(|\mathbf{P}|)$, which yields
\begin{equation*} 
\mathbf {J}_{P}=\mathrm {E}_{\underline {\mathbf {h}}}\left[\frac {\partial ^{2}\mathcal {L}\left ({\mathbf {\underline {h}};{\mathbf {P}}}\right)}{\partial \mathbf {\underline {h}}\partial \mathbf {\underline {h}}^{H}}\right]={\mathbf {P}}^{-1}.
\end{equation*}
Using $\mathcal {L}\left({\mathbf {\underline {y}}|\mathbf {\underline {h}};{\mathbf {P}}}\right)=\frac{1}{\sigma_v^2} ||\mathbf{y}-\widetilde{\mathbf{X}}\mathbf{h}||^2$, the $\mathbf{J}_D$ is
\begin{equation*} 
\mathbf{J}_{D} = \mathrm {E}_{\left ({\underline {\mathbf {y}},\underline {\mathbf {h}}}\right)} \left[\frac {\partial ^{2}\mathcal {L}\left ({\mathbf {\underline {y}}|\mathbf {\underline {h}};{\mathbf {P}}}\right)}{\partial \mathbf {\underline {h}}\partial \mathbf {\underline {h}}^{H}}\right]=\sigma ^{-2}\bar{\mathbf{X}}^{H}\bar{\mathbf{X}},
\end{equation*}
with the BCRB given by
\begin{equation*}
\mathbf {J}_{B}^{-1}=\left ({\sigma ^{-2}\widetilde{\mathbf {X}}^{H}\widetilde {\mathbf {X}}+{\mathbf {P}}^{-1} }\right)^{-1}.
\end{equation*}
After substituting ${\mathbf {P}}=\mathbf{\Psi} \otimes \mathbf{I}_K$ in the above equation, the BCRB bound for $\widehat{\mathbf{H}}$ is obtained as\par\noindent\small
\begin{align*} 
\mathrm {E}\left[\hspace {-0.1cm}\left \|{ \widehat {\underline {\mathbf {H}}}-\underline {\mathbf {H}} }\right \|_{F}^{2}\right]\geq& \mathrm{Tr}\left[\hspace {-0.1cm}\left ({\sigma ^{-2}\left (\widetilde{\mathbf{X}}^{H}\widetilde{\mathbf{X}}\right)\otimes \mathbf {I}_{K}+\boldsymbol {\Psi }^{-1}\otimes \mathbf {I}_{K}}\right)^{-1}\right].
\end{align*}\normalsize
\end{IEEEproof}
In order to derive the BCRBs for the scenarios with group, joint, and joint-cum-group sparse CCIR matrix estimation one can replace the hyperparameter covariance matrix $\boldsymbol {\Psi }$ above by the corresponding hyperparameter covariance matrix in the above expression. Next, in order to show the efficacy of the proposed algorithm we have carried out comprehensive simulations on the simulated as well as practical CCIR obtained by RFView®.
\section{Numerical Experiments}
\label{sec:IV}
\begin{figure}
    \centering
    \includegraphics[width=\linewidth]{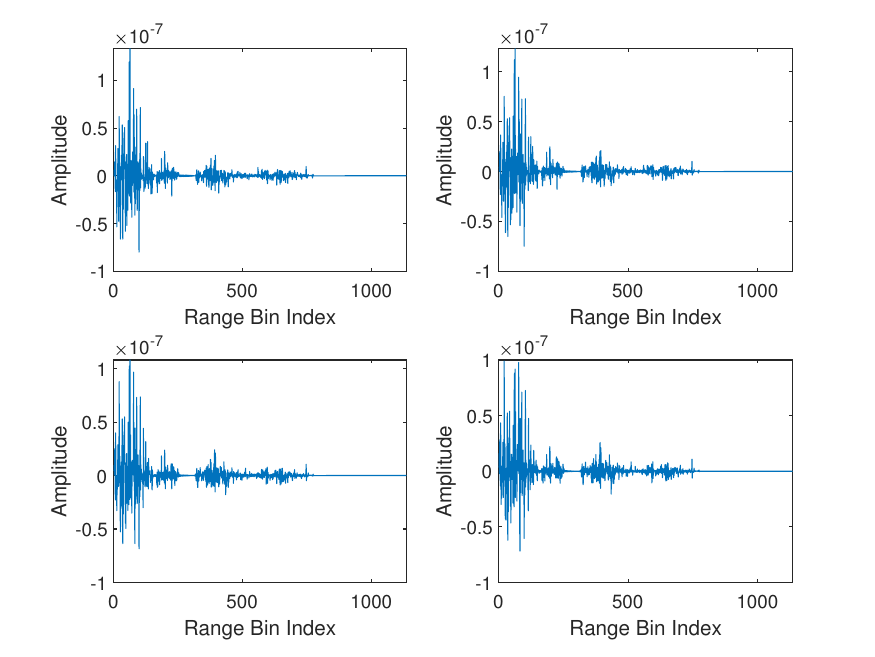}
    \caption{Different channels obtained from RFview depicting the sparsity in the range bin domain.}
    \label{fig:RFView}
\end{figure}
We assessed the efficacy of the covariance-free Bayesian learning techniques for simultaneous, joint, and joint-cum-group sparse MIMO clutter CIR estimation using synthetic data generated from RFView® \cite{RFView}, an advanced modeling and simulation software. {The RFView® Modeling \& Simulation tool, developed by Information Systems Laboratories, Inc. (ISL), is an advanced, site-specific RF simulation and analysis tool. It offers highly precise characterization of complex RF environments, making it ideal for system analysis, test planning, generating high-fidelity synthetic data, and developing signal processing algorithms \cite{gogineni2022high,RFView}.} RFView® intelligently divides the clutter region into separate patches based on publicly available topographical data and land cover classifications to accurately model ground clutter returns for radio frequency systems. The dataset produced by RFView® represents a monostatic radar platform moving at a velocity of $100$ m/s across Southern California. In our evaluation, we considered a MIMO scenario featuring $N=3$ transmitters and $M=2$ receivers. Each transmitter emitted $K=64$ pulses of a Linear Frequency Modulated (LFM) waveforms with a bandwidth of $100$, $200$ and $300$ KHz, respectively, and all of them have pulse repetition frequency of $1$ KHz. The CCIR generated by RFView® has dimensions $3 \times 2 \times 64 \times 1134$, where $R=1134$ denotes the number of range bins.\par 
\begin{figure}
    \centering
    \includegraphics[width=\linewidth]{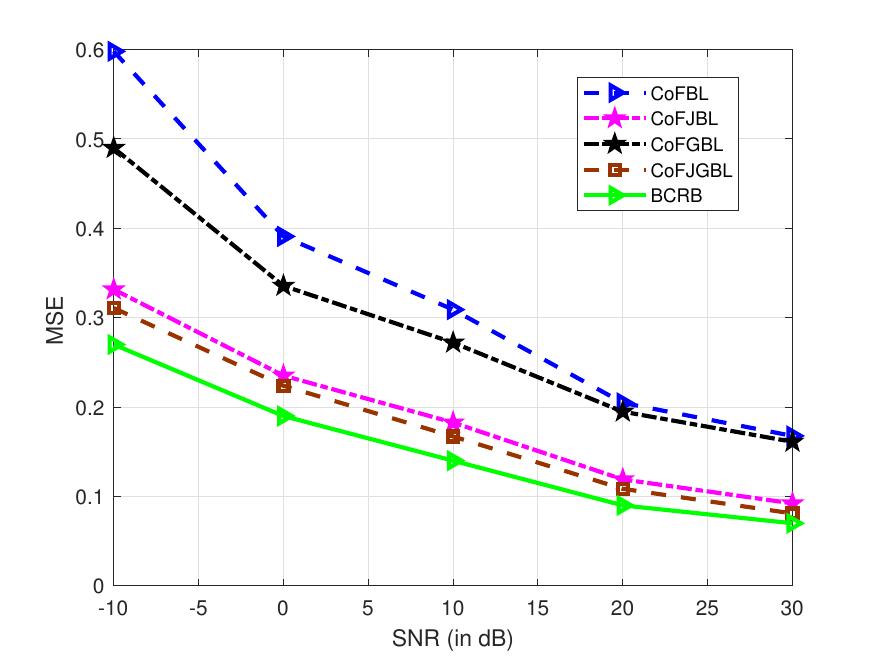}
    \caption{MSE versus SNR for the proposed sparse CCIR estimation algorithms.}
    \label{fig:2}
\end{figure}
Fig. \ref{fig:2} depicts the MSE performance of the various proposed MSBL-based sparse CCIR estimation methods as a function of signal-to-noise ratio (SNR). As expected, with increasing SNR, the MSE tends to decrease for all the proposed algorithms because, with an increment in SNR, the distortion due to noise tends to decrease, which results in improved estimation performance. It is also evident from the figure that joint, group, and joint-cum-group BL algorithms not only reduce the number of hyperparameters but also yield better MSE performance in comparison to the CoFBL algorithm in the same order as mentioned above. This is happening because we are able to exploit the sparsity structure efficiently.\par

\begin{figure}
    \centering
    \includegraphics[width=\linewidth]{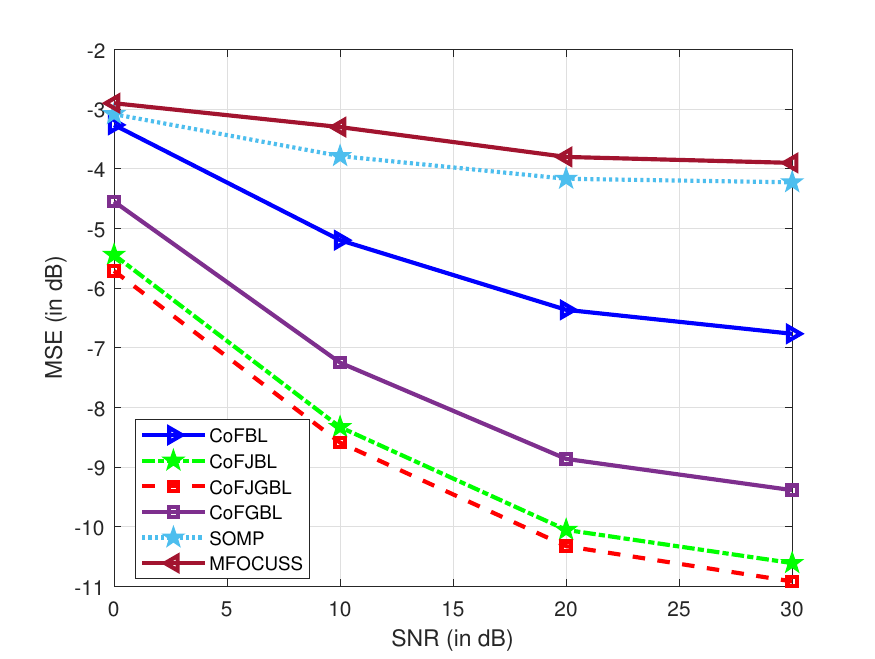}
    \caption{MSE versus SNR performance comparison for the proposed algorithms and the SOMP and MFOCUSS algorithms.}
    \label{fig:3}
\end{figure}
{Fig. \ref{fig:3} demonstrates the performance comparison with other existing simultaneous sparse signal estimation methods in the literature namely simultaneous orthogonal matching pursuit (SOMP)  \cite{TROPP2006572} and \textit{m}ultiple \textit{foc}al \textit{u}nderdetermined \textit{s}ystem \textit{s}olver (MFOCUSS) \cite{1453780}. Herein, the considered algorithms solve equation (14) and their respective MSEs are plotted.} It can be readily seen that the proposed Bayesian learning algorithm outperforms the SOMP and MFOCUSS algorithms which are sensitive to the dictionary matrix and require prior information about the sparsity level. {The poor performance of the SOMP can be attributed to its sensitivity to the stopping parameter as well as to the sensing matrix $\widetilde{\mathbf{X}}$ \cite{990897}, while the poor performance of the MFOCUSS arises due to its convergence deficiencies and sensitivity to the regularization parameter \cite{558475}. By contrast, the proposed CoFBL-based design is robust to the sensing matrix $\widetilde{\mathbf{X}}$, and the well-established properties of its cost-function of (19),
as described in \cite{1315936,4244754}, followed by the EM-framework, guarantees convergence to sparse solutions.}

\begin{figure}
    \centering
    \includegraphics[width=\linewidth]{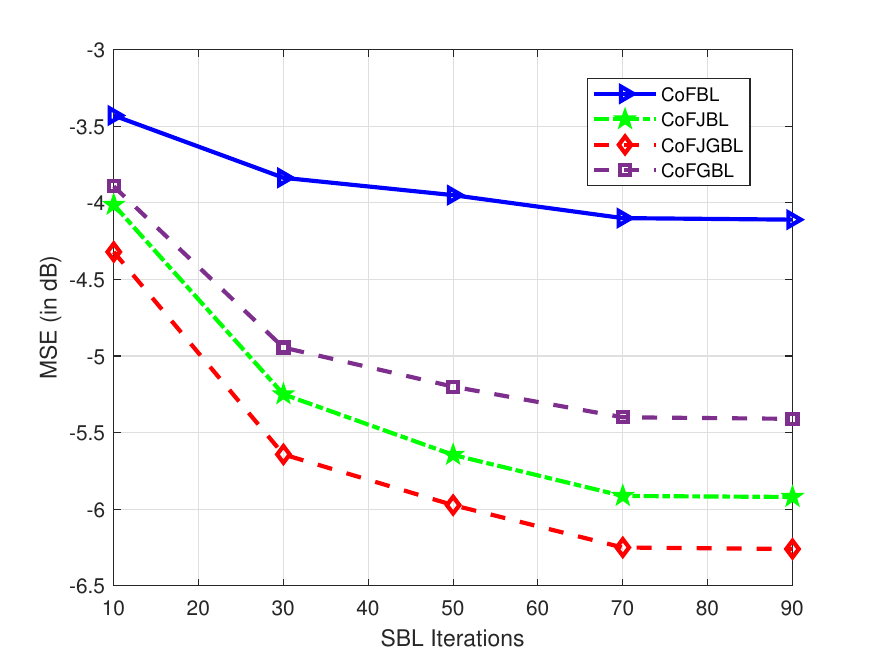}
    \caption{MSE versus EM iterations for the proposed algorithms.}
    \label{fig:4}
\end{figure}
\begin{figure}
    \centering
    \includegraphics[width=\linewidth]{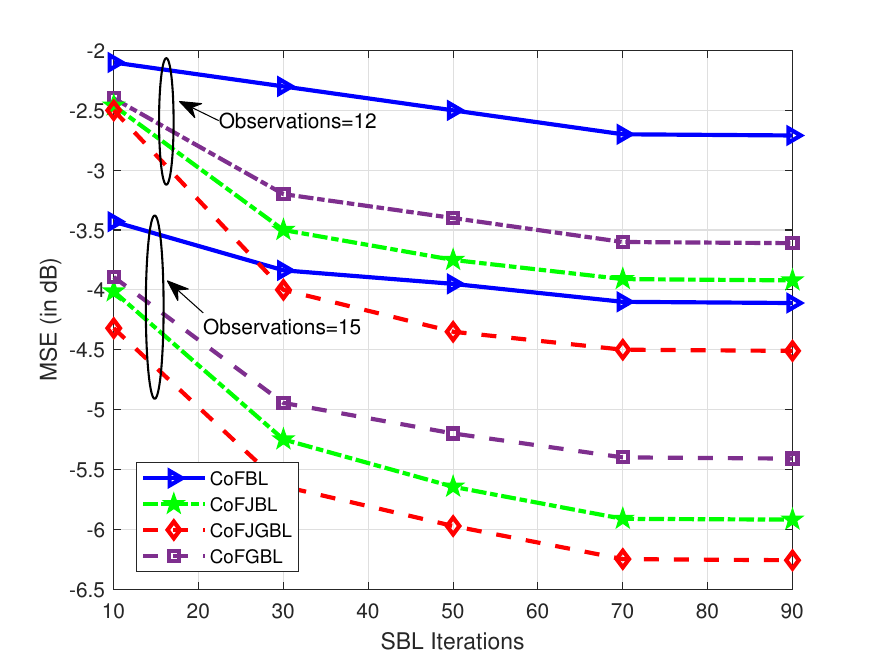}
    \caption{MSE versus EM iterations for the proposed algorithms for different observations.}
    \label{fig:5}
\end{figure}

Next, two plots are derived for the simulated data. Fig. \ref{fig:4} plots MSE as a function of a number of EM iterations. As the number of EM iterations increases the hyperparameters are achieving convergence and the proposed design is able to capture the sparsity in a better way. Fig. \ref{fig:5} plots the MSE as a function of SBL or EM iterations for the different number of measurements at the receiver. It can be readily observed from the figure as we reduce the number of observations the MSE performance deteriorates because there is very little information available for the underlying quantity to be estimated.

{
\begin{table*}
\centering
\caption{Accuracy versus efficiency trade off between the conventional EM and proposed CoFBL algorithms}\label{Table_accuracy}
\begin{tabular}{||l|c|c|c|c|}
\hline
EM-iteration &  \multicolumn{2}{|c|}{Conventional SBL} & \multicolumn{2}{|c|}{CoFBL} \\
\hline
& Accuracy ($\%$) & Run Time (Seconds) & Accuracy ($\%$) & Run Time (Seconds) \\
\hline
10 & 51.58 & 3.91 & 50.26 & 1.4\\
\hline
20   & 21.05 &  7.31  & 16.96 & 2.21\\
\hline
30 & 4.38  &  11.42   & 3.65 & 3.05 \\
\hline
40 & 3.29  &   15.89 & 3.20 & 4.10 \\
\hline
50 & 2.94 &   19.65  & 2.91 & 5.04 \\
\hline
\end{tabular}
\end{table*}}
{Table-\ref{Table_accuracy} demonstrates the accuracy and efficiency as the normalized MSE (in percentage) and Python run time (in seconds), respectively. We have used Intel core i9 2.5 GHz processor 32 GB RAM to carry out this numerical analysis. It has been observed that the proposed CoFBL algorithm has better estimation accuracy and also a very efficient run-time. This shows the effectiveness of the proposed algorithm.}

\begin{figure}
    \centering
    \includegraphics[width=\linewidth]{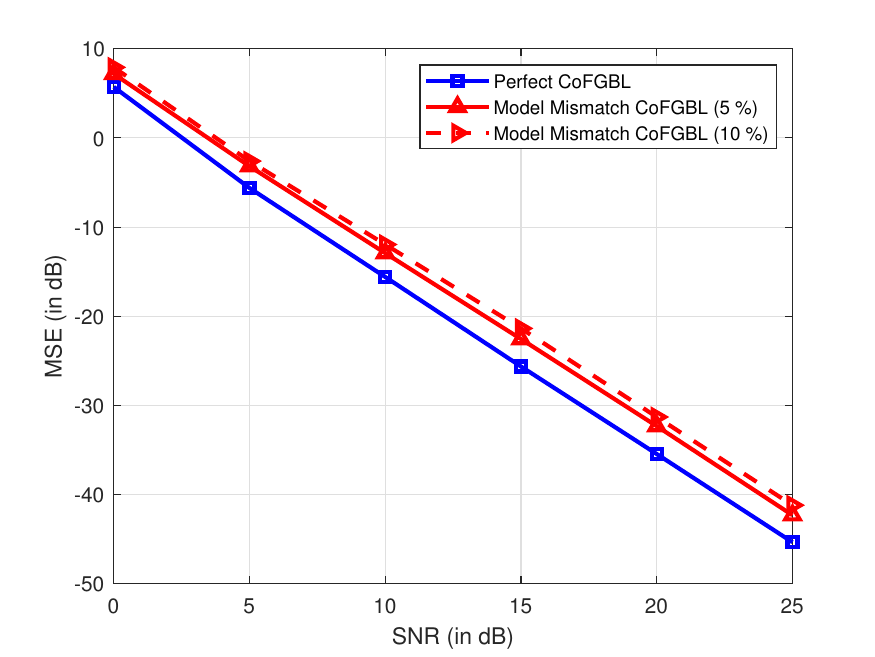}
    \caption{Model mismatch analysis}
    \label{fig:mod_mismatch}
\end{figure}

{Fig. \ref{fig:mod_mismatch} showcases the MSE performance of the CoFGBL algorithm in different scenarios, considering the SNR as a variable. The first scenario explores the case where the underlying CCIR adheres to the group sparsity assumption. In the second and third scenarios, a model mismatch of 5\% and 10\% is introduced, respectively. As anticipated, when the actual sparsity deviates from the group sparsity assumption, the MSE performance of the CoFGBL algorithm is adversely affected, resulting in a higher MSE compared to the ideal scenario that conforms to the group sparse structure. At low SNR levels, the impact on MSE is not significant due to the dominant effect of noise. However, as the SNR increases, the difference in MSE becomes more noticeable, with the magnitude of mismatch directly influencing the observed increase.} \par
\begin{figure}
    \centering    
    \includegraphics[width=\linewidth]{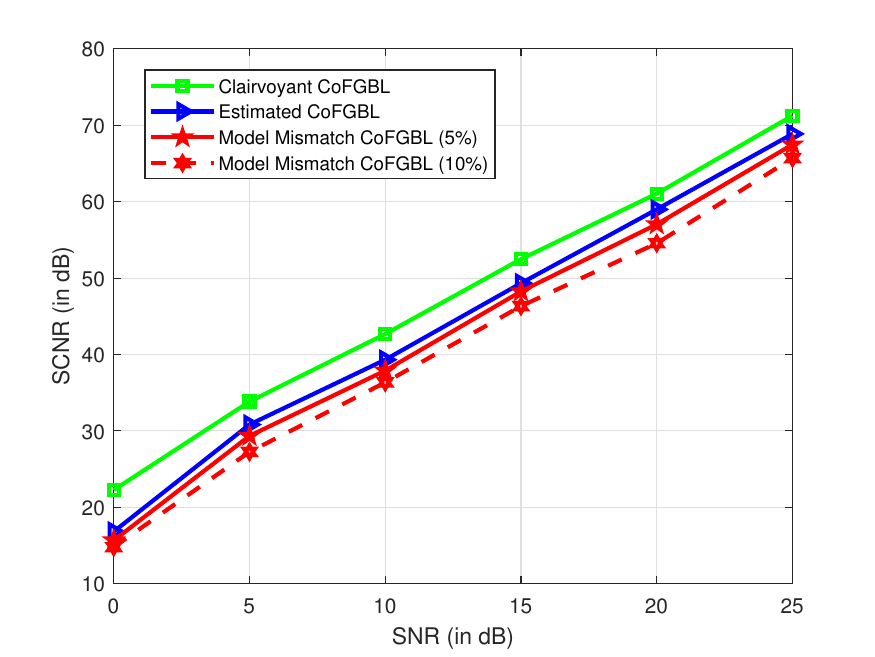}
    \caption{SCNR analysis for the CoFGBL algorithm in model mismatch scenario with MVDR.}
    \label{fig:COFGBL_SNR}
\end{figure}

{Further, we demonstrate the effect of  model mismatch on CoFAR's target detection performance. We considered a scenario for a target moving with a Doppler velocity of 100 km/h. The received signal is
 \begin{align}
   \mathbf{y}=\mathbf{H}_c \mathbf{x}+\mathbf{H}_t \left(\mathbf{x} \odot \mathbf{p} \right)+ \mathbf{n},
 \end{align}
where $\mathbf{H}_c$, $\mathbf{H}_t$, $\mathbf{x}$, $\mathbf{p}$, and $\mathbf{n}$ represent the convolution CCIR matrix, convolution target matrix, transmit waveform vector, Doppler vector, and receive noise vector, respectively. After estimating CCIR using the proposed CoFGBL algorithm, we evaluate the minimum variance distortionless response (MVDR) beamformer \cite{1449208} of the clutter channel for three scenarios: true CCIR (clairvoyant), CoFGBL-estimated CCIR with known model (estimated CoFGBL), and CoFGBL-estimated CCIR with model mismatch, respectively. Two model mismatch scenarios are considered, wherein $5$ and $10$ percent additional non-zero elements are added at random locations that were not included in the group sparse structure. For each of these MVDR beamformers, we  evaluate the respective signal-to-clutter-plus-noise ratio (SCNR). It follows from Fig. \ref{fig:COFGBL_SNR} that the  SCNR performance of estimated CoFGBL algorithm is closest to the clairvoyant case, while the scenarios with model mismatch results in SCNR loss. For instance, at SNR $=10$ dB, the SCNR degrades by $\approx1$ and $\approx2$ dB below estimated CoFGBL performance for $5$ and $10$ percent mismatch, respectively.}

\begin{figure}
    \centering
    \includegraphics[width=\linewidth]{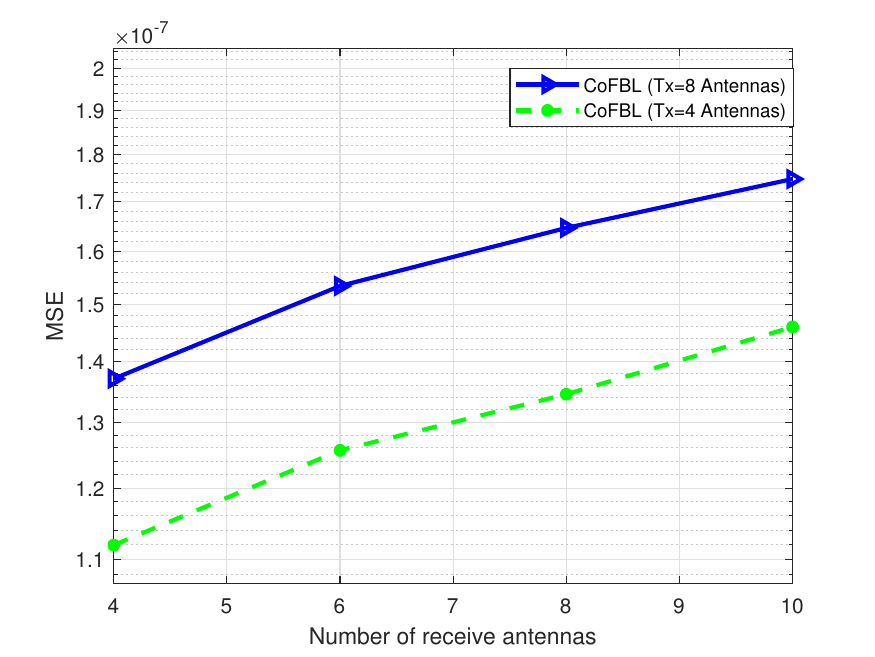}
    \caption{MSE as function of receive antennas.}
    \label{fig:TX_RX}
\end{figure}

{Fig. \ref{fig:TX_RX} shows the MSE performance of the proposed CoFBL technique for the different number for different  receive antennas while the transmit antennas are set equal to $4$ and $8$. Since, increasing the number of antennas will increase the dimension of the unknown sparse CCIR vector $\mathbf{h}_k \in \mathbb{C}^{NMR}$, where $N$, $M$, and $R$ represents the number of transmit antennas, receive antennas, and range bins, respectively. Hence the MSE increases but as it can be seen in the figure that it is not a substantial increment. Hence, one can conclude that the estimation performance is not getting affected much with the increase in number of transmit and receive antennas. On the other hand, this would increase the computation complexity of conventional SBL algorithm, i.e.. $\mathcal{O}\left(N^3M^3R^3\right)$. While, the computational complexity of CoFBL algorithm is $\mathcal{O}\left(UL\tau\right)$, which is linear in terms of $N$ and $M$. Hence, for a practical MIMO radar with more number of transmit and receive antennas the CoFBL becomes more computationally effective.}

\begin{figure}
    \centering
    \includegraphics[width=\linewidth]{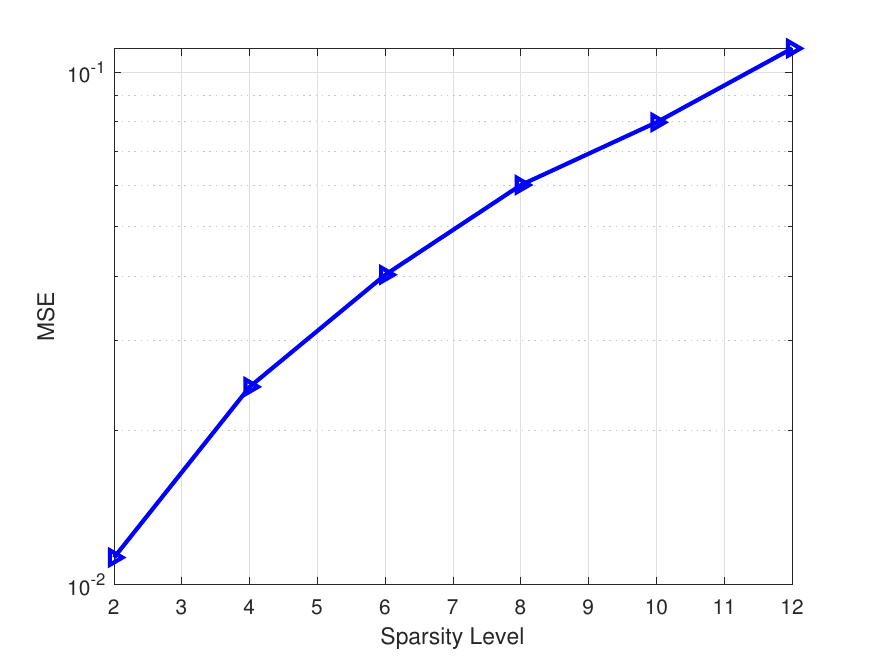}
    \caption{MSE as a function of sparsity level.}
    \label{fig:sparsity}
\end{figure}
\begin{figure}
    \centering
    \includegraphics[width=\linewidth]{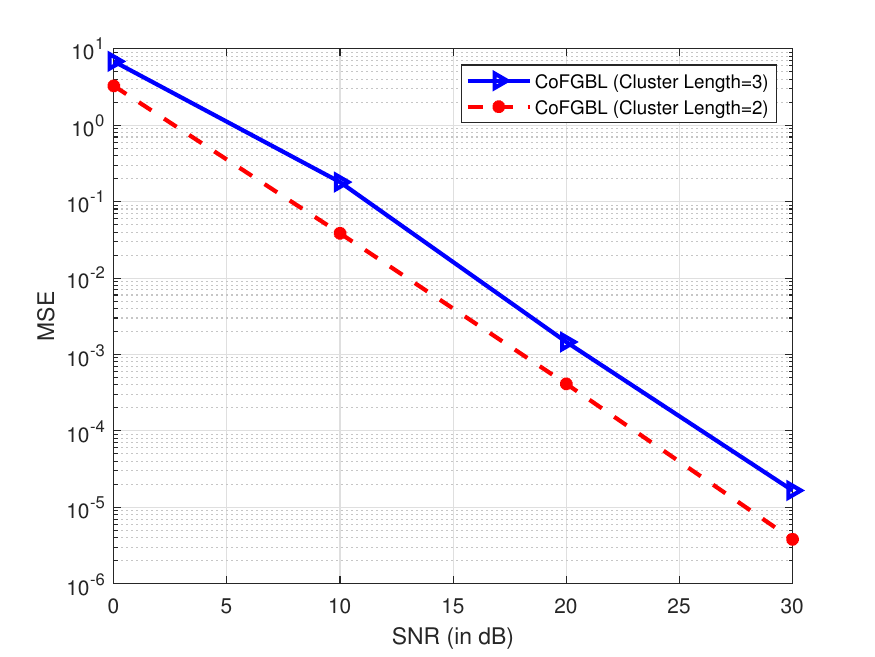}
    \caption{MSE versus SNR for CoFGBL with different cluster length}
    \label{fig:CoFGBL_CL}
\end{figure}
{Fig. \ref{fig:sparsity} demonstrates the MSE of the CoFBL algorithms as a function of sparsity level of the sparse CCIR matrix, i.e., the number of non-zero rows. It can be readily observed that as the number of non-zero rows increases the MSE increases since, the number of elements which are to be estimated also increases and hence the estimation error increases. Fig. \ref{fig:CoFGBL_CL} depicts MSE performance of the CoFGBL algorithm for the different cluster lengths as a funtion of SNR. As it can be readily seen in the figure that with increasing cluster length one will have the lower computational complexity but it comes at the cost of increased MSE.} \par
\begin{figure}
    \centering
    \includegraphics[width=\linewidth]{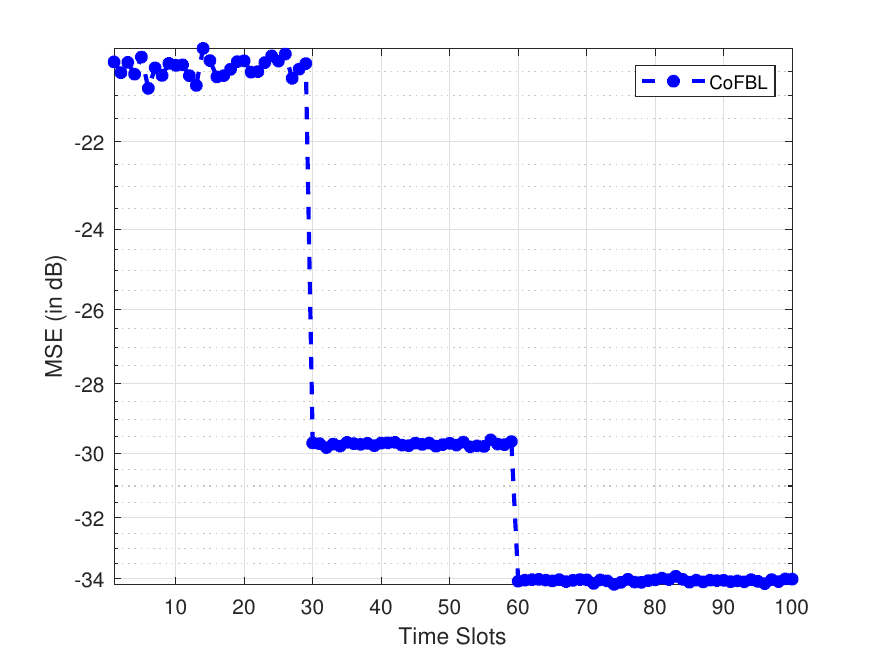}
    \caption{CoFBL performance performance with changing environment.}
    \label{fig:COFBL_Chan_Time}
\end{figure}
{Fig. \ref{fig:COFBL_Chan_Time} illustrates the MSE performance of the proposed CoFBL algorithm in a dynamic environment. It depicts the algorithm's ability to adapt to changing conditions over time. In this simulation, the sparsity profile remains constant for the initial 30 time slots. From the 31st to the 60th time slot, there is a 10 percent decrease in the sparsity profile, followed by an additional 10 percent decrement from the 60th to the 100th time slot. This clearly demonstrates that the CoFBL algorithm successfully tracks the changing environment. As the number of elements to be estimated decreases, there is a corresponding reduction in the MSE.}

\section{Summary}
\label{sec:V}
In this paper, we analyzed the MIMO CCIR estimation problem in CoFAR. We devised different covariance-free BL-based CCIR estimation approaches that leverage the simultaneous, joint, and joint-cum-group sparsity of the CCIR. The numerical results imply that the suggested BL-based CCIR estimation strategy outperforms the other sparse estimation techniques such as SOMP and MFOCUSS while using fewer resources such as pilot length. This implies that using the proposed algorithms one can save critical resources like time and power, which can be further utilized for the primary function of target detection in CoFAR. 

{In the future, the proposed CoFBL technique and its variants show potential for utilization in conventional radar applications such as target imaging and parameter estimation \cite{8529199, 8537983}. These applications are characterized by scenarios where only a few targets are present, creating an ideal condition that can be modeled as a sparse recovery problem. Additionally, the integration of sensing and communication is an emerging research area for 6G wireless systems \cite{mishra2024signal,8954710,9898900}, where both radar and communication systems operate in the same frequency band. 6G technologies, including terahertz/millimeter-wave (mmWave) and orthogonal time-frequency space (OTFS), exhibit sparsity in the wireless channels. When combined with radar systems that encounter only a limited number of targets or clutter patches, the problem can once again be formulated as a sparse recovery problem that can be effectively addressed using the CoFBL technique.} 

Sea clutter \cite{bringi2024advances1,bringi2024advances3} presents unique modeling complexities due to its non-stationary nature. Accurately representing it with the models considered in this paper requires thorough investigation. Preliminary work in this area can be found in \cite{gogineni2022high}. While the current models can handle non-stationary scenarios using the single measurement vector setup, applying this approach to sea clutter remains a topic for future research.

{The CoFBL technique holds promise for various research problems, including joint CCIR estimation and waveform design, and it would be interesting to extend it to scenarios where the underlying sparse CCIR vectors exhibit correlation \cite{kang2021}. Apart from radar applications, the CoFBL and its variants find utility in estimating sparse wireless channels in mmWave \cite{8398433}, terahertz \cite{10149498} and orthogonal time frequency space (OTFS) \cite{9184852} wireless communication systems. In the context of wireless communication, where massive machine type communication involves a large number of connected devices, the demand for computationally efficient channel or parameter estimation algorithms like CoFBL becomes crucial \cite{9423649}.}

\appendices
\section{Group sparse CCIR estimation}
The complete likelihood ratio $\mathcal{L}\left(\mathbf{\Psi}_G|\widehat{\mathbf{\Psi}}^{(l-1)}_G\right)$ in E-step is derived as
\begin{equation} 
\mathbb {E} \bigg [ \log \left [{p\left ({\widetilde {\mathbf {Y}} | {\mathbf {H}} }\right) }\right] \bigg ] + \mathbb {E} \bigg [ \log \left [{ p\left ({ {\mathbf {H}}; \boldsymbol{\Psi }_G }\right) }\right] \bigg ]. 
\end{equation}
M-step: Next, one maximizes $\mathcal{L}\left(\mathbf{\Psi}_G|\widehat{\mathbf{\Psi}}^{(l-1)}_G\right)$ with respect to $\boldsymbol{\Psi}_G$ as
\begin{equation} \label{Gamma_G}
\widehat {\boldsymbol \Psi }^{(l)}_G=\arg \max _{\boldsymbol{\Psi}_G } \;\mathbb {E} \bigg [\log \left [{ p\left ({ {\mathbf {H}}; \boldsymbol{\Psi }_G }\right) }\right] \bigg ], 
\end{equation}
where we don't consider the term independent of $\boldsymbol{\Psi}_G$. Using \eqref{Prior_H2}, substituting $ \log \left [{ p\left ({ {\mathbf {H}}; \boldsymbol{\Psi }_G }\right) }\right] $, one obtains
\begin{equation*}
\widehat {\boldsymbol \Psi }^{(l)}_G \!\equiv \!\arg \max _{\boldsymbol{\Psi}_G } \sum \limits _{i=1}^{\frac{NMR}{d}} \left[{\! -\!dK\log (\bar{\psi}_{i}) \!-\!\! \sum _{k=1}^{K}\! \displaystyle \frac {\mathbb {E} \big \{ \vert \mathbf {h}_{n,m,k}(k) \vert ^{2} \big \}}{\bar{\psi}_{i}} \!}\right]\!,
\end{equation*}
where we ignore the constant terms in the optimization that are independent of $\boldsymbol{\Psi}_G$. From the above equation, it can be readily observed that the optimization problem can be decoupled in each $\widehat {\bar{\psi}}_{i}$. Hence, the estimates $\widehat {\bar{\psi}}_{i}^{(l)}$ can be evaluated using
\begin{equation}\label{Gamma_GMSBL}
\widehat {\bar{\psi}}_{i}^{(l)}= \frac {1}{dK}  \sum _{k=1}^{K}\mathbb {E}_{\mathbf {H} \vert {\mathbf {Y}}; \widehat{\boldsymbol{\Psi}}_G^{(l-1)}} \bigg[ \vert \mathbf {h}_{n,m,k}(k) \vert ^{2} \bigg]. 
\end{equation}
The conditional expectation $\mathbb {E}_{ {\mathbf {H}} \vert {\mathbf {Y}}}[.]$ can be evaluated using \textit{a posteriori} pdf $p({\mathbf {H}} \vert {\mathbf {Y}};\boldsymbol{\Psi }^{(l-1)}_G)$ of $\mathbf {H}$, and is defined as $p\left ({{\mathbf {H}}|{\mathbf {Y}}}; \widehat {\boldsymbol{\Psi }}^{(l-1)}_G\right) = \mathcal {CN}\left ({\boldsymbol {\mathcal {M}}^{(l)}_G, {\boldsymbol{\Sigma }}^{(l)}_G}\right)$. The quantities ${\boldsymbol{\Sigma }}^{(l)}_G \in \mathbb{C}^{CGNt \times GNt}$ and $\boldsymbol {\mathcal {M}}^{(l)}_G \in \mathbb{C}^{CGNt \times GNt}$ are derived by substituting $\boldsymbol{\Psi }={\boldsymbol{\Psi }}^{(l-1)}_G$ in Eq. \eqref{Covariance}, and ${\boldsymbol{\Sigma }}={\boldsymbol{\Sigma }}^{(l)}_G$ in Eq. \eqref{Mean}, respectively. Hence, $\mathbb {E} \bigg[\vert \mathbf {h}_{n,m,k}(i) \vert ^{2} \bigg]$ can be evaluated as
\begin{align*}
&\hspace {-.5pc} \mathbb {E} \bigg[\vert \mathbf {h}_{n,m,k}(i) \vert ^{2} \bigg] = \left \vert{ \boldsymbol {\mathcal {M}}^{(l)}_G\big [\tilde{m},k \big] }\right \vert ^{2} +\, {\boldsymbol{\Sigma }}^{(l)}_G \big [\tilde{m},\tilde{m} \big],
\end{align*}
which upon substituting into \eqref{Gamma_GMSBL}, yields the desired hyperparameter update of \eqref{Para_Est_GMSBL}.
\section{Joint sparse CCIR estimation}
For the scenario with joint sparsity, the complete likelihood ratio $\mathcal{L}\left(\mathbf{\Psi}_J|\mathbf{\Psi}_J^{(l-1)}\right)$ in the E-step is
\begin{equation} 
\mathbb {E} \bigg[ \log \left [{p\left ({\widetilde {\mathbf {Y}} | {\mathbf {H}} }\right) }\right] \bigg]+ \mathbb {E} \bigg[  \log \left [{ p\left ({ {\mathbf {H}}; \boldsymbol{\Psi }_J }\right) }\right] \bigg]. 
\end{equation}
In M-step, one maximizes $\mathcal{L}\left(\mathbf{\Psi}_J|\mathbf{\Psi}_J^{(l-1)}\right)$ with respect to $\boldsymbol{\Psi}$ as
\begin{equation} 
\widehat {\boldsymbol \Psi }_J^{(l)}=\arg \max _{\boldsymbol{\Psi}_J} \;\mathbb {E} \bigg[ \log \left [{ p\left ({ {\mathbf {H}}; \boldsymbol{\Psi}_J }\right) }\right] \bigg], 
\end{equation}
ignoring the terms independent of $\boldsymbol{\Psi}_J$. Furthermore, using  \eqref{Prior_JSBL}, and leaving out the terms independent of $\boldsymbol{\Psi}_J$, the hyperparameter covariance matrix estimate is\par\noindent\small
\begin{align*}
&\widehat {\boldsymbol{\Psi}}_J^{(l)} \!\equiv \!\arg \max _{\boldsymbol{\Psi}_J } \sum \limits _{i=1}^{R} \left[{\! -\!NMK\log (\tilde{\psi}_{i}) - 
\sum _{n=1,m=1,k=1}^{N,M,K}\! \displaystyle \frac {\mathbb {E} \big \{ \vert \mathbf {h}_{n,m,k}(i) \vert ^{2} \big \}}{\tilde{\psi}_{i}} \!}\right].
\end{align*}\normalsize
Similar to previous group sparse scenario, the hyperparameter estimates $\widehat {\tilde{\psi} }_{i}^{(l)}$ can be evaluated using
\begin{equation}\label{Gamma_JMSBL}
\widehat {\tilde{\psi} }_{i}^{(l)}= \frac {1}{NMK} \sum _{n=1}^{N} \sum _{m=1}^{M} \sum _{k=1}^{K}\mathbb {E}_{{\mathbf {H}} \vert {\mathbf {Y}}; \widehat{\boldsymbol{\Psi}}_J^{(l-1)}} \bigg[\vert \mathbf {h}_{n,m,k}(i) \vert ^{2} \bigg]. 
\end{equation}
The \textit{a posteriori} pdf $p\left ({{\mathbf {H}}|{\mathbf {Y}}}; \widehat {\boldsymbol{\Psi }}_J^{(l-1)}\right) = \mathcal {CN}\left ({\boldsymbol {\mathcal {M}}_J^{(l)}, {\boldsymbol{\Sigma }}_J^{(l)}}\right)$ is used to evaluate condition expectation $\mathbb {E}_{ {\mathbf {H}} \vert {\mathbf {Y}}}[.]$. Hence, $\mathbb {E} \bigg [\vert \mathbf {h}_{n,m,k}(i) \vert ^{2} \bigg]$ can be evaluated as
\begin{align*}
&\hspace {-.5pc} \mathbb {E} \bigg [\vert \mathbf {h}_{n,m,k}(i) \vert ^{2} \bigg] = \frac {1}{NMK} \sum _{n=1}^{N} \sum _{m=1}^{M} \sum _{k=1}^{K}\left \vert{ \boldsymbol {\mathcal {M}}_J^{(l)}\big [\bar{m},k \big] }\right \vert ^{2} \\&+\, \widetilde {\boldsymbol{\Sigma }}_J^{(l)} \big [\bar{m},\bar{m} \big],
\end{align*}
which upon substituting into \eqref{Gamma_JMSBL}, yields the desired hyperparameter update of \eqref{Para_Est_JMSBL}.

\section{Joint-cum-group sparse CCIR estimation}
For the scenario with joint-cum-group sparsity, in E-step one evaluates $\mathcal{L}\left(\mathbf{\Psi}_{JG}|\mathbf{\Psi}_{JG}^{(l-1)}\right)$ using
\begin{equation} 
\mathbb {E} \bigg[ \log \left [{p\left ({\widetilde {\mathbf {Y}} | {\mathbf {H}} }\right) }\right] \bigg] + \mathbb {E} \bigg[\log \left [{ p\left ({ {\mathbf {H}}; \boldsymbol{\Psi }_{JG} }\right) }\right] \bigg]. 
\end{equation}
Furthermore, in M-step, one maximizes $\mathcal{L}\left(\mathbf{\Psi}_{JG}|\mathbf{\Psi}_{JG}^{(l-1)}\right)$ with respect to $\boldsymbol{\Psi}$ which can be mathematically represented as
\begin{equation} 
\widehat {\boldsymbol \Psi }_{JG}^{(l)}=\arg \max _{\boldsymbol{\Psi}_{JG} } \;\mathbb {E} \bigg[ \log \left [{ p\left ({ {\mathbf {H}}; \boldsymbol{\Psi }_{JG} }\right) }\right] \bigg], 
\end{equation}
Furthermore, using \eqref{Prior_JGSBL}, one can substitute $\log \left[{ p \left({ {\mathbf {H}}; \boldsymbol{\Psi}_{JG} }\right) }\right]$ above, one obtains\par\noindent\small
\begin{align}\label{Gamma_JGMSBL}
\widehat {\boldsymbol \Psi }_{JG}^{(l)} &\!\equiv \!\arg \max _{\boldsymbol{\Psi}_{JG} } \sum \limits _{i=1}^{R/n} \left[{\! -\!NMK\log (\underline{\psi}_{i}) 
\! -\!\! \sum _{n=1,m=1,k=1}^{N,M,K}\! \displaystyle \frac {\mathbb {E} \big \{ \vert \mathbf {h}_{n,m,k}(i) \vert ^{2} \big \}}{\underline{\psi}_{i}} \!}\right]\!.
\end{align}\normalsize
Finally, the hyperparameter estimates $\widehat {\underline{\psi} }_{i}^{(l)}$ can be evaluated using
\begin{equation*}
\widehat {\underline{\psi} }_{i}^{(l)}= \frac {1}{nNMK} \sum _{n=1}^{N} \sum _{m=1}^{M} \sum _{k=1}^{K}\mathbb {E}_{{\mathbf {H}} \vert {\mathbf {Y}}; \widehat{\boldsymbol{\Psi}}_{JG}^{(l-1)}} \bigg[ \vert \mathbf {h}_{n,m,k}(i) \vert ^{2} \bigg]. 
\end{equation*}
The conditional expectation $\mathbb {E}_{ {\mathbf {H}} \vert {\mathbf {Y}}}[.]$can be evaluating using \textit{a posteriori} pdf $p\left ({{\mathbf {H}}|{\mathbf {Y}}}; \widehat {\boldsymbol{\Psi }}_{JG}^{(l-1)}\right) = \mathcal {CN}\left ({\boldsymbol {\mathcal {M}}_{JG}^{(l)}, {\boldsymbol{\Sigma }}_{JG}^{(l)}}\right)$. The matrices  ${\boldsymbol{\Sigma }}_{JG}^{(l)} \in \mathbb{C}^{CGNt \times GNt}$ and $\boldsymbol {\mathcal {M}}_{JG}^{(l)} \in \mathbb{C}^{CGNt \times GNt}$ are derived by replacing $\boldsymbol{\Psi }={\boldsymbol{\Psi }}_{JG}^{(l-1)}$ in Eq. \eqref{Covariance}, and ${\boldsymbol{\Sigma }}={\boldsymbol{\Sigma }}_{JG}^{(l)}$ in Eq. \eqref{Mean}, respectively. Finally, one obtains
\begin{align*}
&\hspace {-.5pc} \mathbb {E} \bigg [ \vert \mathbf {h}_{n,m,k}(i) \vert ^{2} \bigg] = \frac {1}{nNMK} \sum _{n=1}^{N} \sum _{m=1}^{M} \sum _{k=1}^{K}\left \vert{ \boldsymbol {\mathcal {M}}_{JG}^{(l)}\big [\underline{m},k \big] }\right \vert ^{2} \\&+\, \frac {1}{nNM} \sum _{n=1}^{N} \sum _{m=1}^{M}\widetilde {\boldsymbol{\Sigma }}_{JG}^{(l)} \big [\underline{m},\underline{m} \big],
\end{align*}
which upon substituting into \eqref{Gamma_JGMSBL}, yields the desired hyperparameter update of \eqref{Para_Est_JGMSBL}.

\bibliographystyle{IEEEtran}
\bibliography{reference}

\end{document}